\documentclass[a4paper,11pt]{article}
\usepackage{jcappub}

\usepackage[nameinlink,noabbrev]{cleveref}
\usepackage{xspace}
\usepackage{url}
\usepackage{aas_macros}
\usepackage{booktabs}
\graphicspath{{Figures/}}
\usepackage{scalerel}
\usepackage{multirow}

\usepackage{orcidlink}

\title{Primordial Features in light of the Effective Field Theory of Large-Scale Structure}


\newcommand{\planck}{\emph{Planck}}

\newcommand{\lcdm}{$\Lambda$CDM}

\newcommand{\Omo}{\ensuremath{\Omega_{\text{m},0}}}

\newcommand{\vect}[1]{\ensuremath{\boldsymbol{#1}}}

\newcommand{\OSpec}{\textit{One Spectrum }}
\newcommand{\classy}{\texttt{class}}

\newcommand{\pybird}{\texttt{PyBird}}
\newcommand{\gd}{\texttt{GetDist}}

\crefname{equation}{Eq.}{Eqs.}
\crefname{section}{Section}{Sections}
\crefname{figure}{Fig.}{Figs.}
\crefname{table}{Table}{Tables}
\crefname{appendix}{Appendix}{Appendices}
\Crefname{figure}{Figure}{Figures}
\Crefname{equation}{Equation}{Equations}
\Crefname{section}{Section}{Sections}
\Crefname{table}{Table}{Tables}

\definecolor{llgray}{gray}{0.93}
\definecolor{lgray}{gray}{0.83}
\definecolor{deepmagenta}{rgb}{0.8, 0.0, 0.8}
\definecolor{ballblue}{rgb}{0.13, 0.67, 0.8}
\definecolor{celestialblue}{rgb}{0.29, 0.59, 0.82}
\definecolor{RedWine}{rgb}{0.743,0,0}
\definecolor{DarkGreen}{rgb}{0,0.6,0}

\usepackage{fontawesome5}
\newcommand{\github}[1]{\href{#1}{\faGithub}
}

\author[a,b]{Rodrigo~Calderón\orcidlink{0000-0002-8215-7292},}
\author[c]{Th\'eo Simon\orcidlink{0000-0001-7858-6441},}
\author[b,d]{Arman~Shafieloo\orcidlink{0000-0001-6815-0337},}
\author[e,f,g]{Dhiraj Kumar Hazra\orcidlink{0000-0001-7041-4143}}

\affiliation[a]{CEICO, Institute of Physics of the Czech Academy of Sciences,\\
Na Slovance 1999/2, 182 21, Prague, Czech Republic}
\affiliation[b]{Korea Astronomy and Space Science Institute,
Daejeon 34055, South Korea}
\affiliation[c]{Laboratoire Univers et Particules de Montpellier (LUPM), 
Centre national de la recherche scientifique (CNRS) et Universit\'e de Montpellier, 34095 Montpellier, France}
\affiliation[d]{University of Science and Technology, 
Daejeon 34113, South Korea}
\affiliation[e]{The Institute of Mathematical Sciences, HBNI, CIT Campus, Chennai 600113, India}
\affiliation[f]{Homi Bhabha National Institute, Training School Complex, Anushakti Nagar, Mumbai 400085, India}
\affiliation[g]{INAF/OAS Bologna, Osservatorio di Astrofisica e Scienza dello Spazio,
Area della ricerca CNR-INAF, via Gobetti 101, I-40129 Bologna, Italy
}

\emailAdd{calderon@fzu.cz}
\emailAdd{theo.simon@umontpellier.fr}

\abstract{
While the simplest inflationary models predict a power-law form of the primordial power spectrum (PPS), various UV complete scenarios predict features on top of the standard power law that leave characteristic imprints in the late-time distribution of matter, encoded in the galaxy power spectrum.
In this work, we assess the validity of the Effective Field Theory of Large Scale Structure (EFTofLSS) and the IR-resummation scheme of \texttt{PyBird} in the context of primordial (oscillatory) features. We find an excellent agreement at the level of the matter power spectrum between N-body simulations and the one-loop EFT predictions, for models commonly studied in the literature.
We then apply the EFTofLSS to the galaxy power spectrum measurements from BOSS LRG and eBOSS QSO to constrain specific global and local features in the PPS. We demonstrate that while such features can improve the fit to cosmic microwave background (CMB) data, they may result in a poorer fit to clustering measurements at low redshift. The resulting constraints on the amplitude of the primordial oscillations are competitive with those obtained from CMB data, despite the well-known damping of oscillations due to non-linear structure formation processes.
For the first time in this context, we jointly analyze the galaxy power spectrum (monopole and quadrupole) in combination with {\it Planck} CMB data to derive strong constraints on the amplitude of primordial features.
This work highlights the EFTofLSS as a powerful tool for testing early universe scenarios on scales that complement CMB observations.

}

\begin{document}
\maketitle
\flushbottom

\section{Introduction}

The standard  model of cosmology (\lcdm) is rooted in the assumption that the Universe experienced  an epoch of accelerated (exponential) expansion during its very early stage, typically known as ``Inflation'' \cite{Starobinsky:1980te,1980ApJ...239..428K,PhysRevD.23.347} (see \cite{Martin:2013tda,Achucarro:2022qrl} for a review). 
A key prediction of the standard inflationary paradigm is the generation of primordial density fluctuations, with a \textit{nearly adiabatic} and \textit{nearly scale-invariant} spectrum. These predictions are in excellent agreement with cosmic microwave background (CMB) observations, spanning a wide range of scales—from the largest, observed by the space-based \planck\ satellite \cite{Planck:2018nkj,Planck:2015sxf,Planck:2018jri}, to the smallest, probed by ground-based experiments such as the Atacama Cosmology Telescope (ACT) \cite{ACT:2020gnv,ACT:2020frw,ACT:2025fju,ACT:2025tim} and the South Pole Telescope (SPT) \cite{SPT-3G:2014dbx,SPT-3G:2021eoc,SPT-3G:2022hvq,SPT-3G:2024atg}.  

Despite the success of the standard inflationary paradigm, numerous alternative scenarios\textemdash often motivated by high-energy physics\textemdash remain consistent with current observations \cite{Planck:2015sxf,Planck:2018jri}. A common prediction of many of these models is the presence of oscillatory features superimposed on the conventional power-law form of the primordial power spectrum (PPS). 
These features have garnered attention for their potential to address certain anomalies in the CMB, such as, the large scale power suppression, localized outliers, the lensing excess ($A_{\rm L}>1$) \cite{Planck:2015bpv,Addison:2015wyg,Motloch:2018pjy} and the apparent preference for a closed universe ($\Omega_k<0$) \cite{Planck:2018vyg,DiValentino:2019qzk} suggested by \planck\ (PR3) data (see \textit{e.g.}, \cite{Hazra:2014jwa, GallegoCadavid:2016wcz,Domenech:2019cyh,Ballardini:2022vzh,Domenech:2020qay}). Furthermore, deviations from the standard slow-roll inflationary dynamics can impact the inferred late-time cosmological parameters, including the Hubble constant and the amplitude of matter fluctuations $S_8\equiv\sigma_8\sqrt{\Omo/0.3}$, helping to alleviate the existing tensions between high and low-redshift estimations of these quantities \cite{Abdalla:2022yfr} when assuming a \lcdm\ expansion history with a power-law spectrum (see \textit{e.g}, \cite{Keeley:2020rmo,Hazra:2022rdl,Stahl:2025qru}).  

From a theoretical perspective, deviations from the nearly adiabatic and scale-invariant spectrum could potentially reveal new physics beyond the standard inflationary framework, and probe fundamental physics at much higher energy scales beyond our current reach.
Current and upcoming measurements of the cosmic microwave background (CMB) and large-scale structure (LSS) offer a powerful means of testing primordial features (see \textit{e.g.}, \cite{Starobinsky:1992ts,PhysRevD.62.043508,Adams:2001vc,Achucarro:2010da,Chen:2011zf,Hu:2014hra,Chluba:2015bqa,Beutler:2019ojk,Ballardini:2019tuc,Slosar:2019gvt,Mergulhao:2023ukp,Terasawa:2025gwa}). A key advantage of LSS data is its complementarity with CMB observations, both in terms of the scales and redshifts they probe. 
Therefore, combining these two probes is essential for breaking degeneracies between parameters governing early universe physics and those shaping late-time evolution. However, at small scales, linear perturbation theory becomes insufficient in predicting the behavior of primordial features in the galaxy overdensity field, as structure formation enters the highly non-linear regime. In particular, these non-linear effects introduce a damping of oscillatory primordial features, as demonstrated by numerical $N$-body simulations \cite{Ballardini:2019tuc,Ballardini:2024dto,Stahl:2025qru}.  

In this paper, we make use for the first time the Effective Field Theory of Large Scale Structure (EFTofLSS)~\cite{Carrasco:2012cv,Baumann:2010tm}, that properly captures the effect of non-linearities coming from the UV physics, in order to constrain features in the PPS. 
This theoretical framework is further accompanied by an IR-resummation scheme which encodes the effects of the long-wavelength displacements on the BAO peaks and the primordial oscillatory features in the matter (and galaxy) power spectrum.
In this paper, we apply this formalism to the monopole and quadrupole of the power spectrum of biased tracers from BOSS DR12 LRG and eBOSS DR16 QSO data in order to obtain constraints on the amplitude of primordial oscillations that are competitive with \planck.
In addition, we exploit the complementary between the LSS and CMB information coming from different scales and redshifts by combining our LSS analysis with the \planck~primary power spectra.
To our knowledge this is the first work that presents a joint analysis of CMB and LSS data, containing both monopole and quadrupole of the galaxy power spectrum, in the context of primordial features. Given the ongoing and upcoming LSS observations (from DESI \cite{DESI:2024hhd}, LSST \cite{LSSTDarkEnergyScience:2012kar} and {\it Euclid} \cite{EUCLID:2011zbd,Euclid:2023shr}) and CMB observations (such as Simons Observatory \cite{SimonsObservatory:2019qwx}, CMB-S4 \cite{CMB-S4:2016ple} and LiteBIRD \cite{Matsumura:2013aja}), our analysis provides a unified framework to obtain stringent constraints on the initial conditions of the Universe with the CMB and LSS.

We consider in this paper two classes of PPS, corresponding to global and local features, that introduce superimposed oscillations on top of the standard power law.
Certain models within these classes, apart from being phenomenologically appealing, are interesting as they can be mapped onto the theoretical parameter space of high-energy theories that predict such primordial features.
We also consider a PPS that has been directly reconstructed from CMB data in order to further gauge the constraining power of LSS data on the initial conditions of the Universe. 
The features selected in this paper are therefore capable of covering a wide variety of models that are either motivated or engineered from the CMB data.

This paper is structured as follows. In \cref{Analysis}, we describe the methods and data used in our analysis. 
In \cref{sec:powerlaw}, we briefly summarize the current LSS and CMB constraints on the vanilla power-law form for the PPS. 
In \cref{sec:Global}, we study \textit{global} feature scenarios (namely, linear and logarithmic primordial oscillaory features). In particular, in \cref{sec:Nbody}, we present the consistency between our EFTofLSS analysis and N-body simulations, before showing our constraints in \cref{sec:global_constraints}.
In \cref{sec:one_spec}, we present our constraints for two (linear and logarithmic) \textit{localized} oscillatory features scenarios.
Finally, in \cref{sec:regMRL}, we study a specific form of the PPS, obtained by deconvolving the observed CMB angular power spectra, before concluding in \cref{sec:conclusion}.

\section{Inference setup}\label{Analysis}

\subsection{Datasets}

In this paper, we perform Markov chain Monte Carlo (MCMC) analyses thanks to the Metropolis-Hasting algorithm implemented in \texttt{Montepython-v3} \github{https://github.com/brinckmann/montepython_public}\cite{Brinckmann:2018cvx,Audren:2012wb}, which is interfaced with the standard \classy~\github{https://github.com/lesgourg/class_public/tree/master}code \cite{Lesgourgues:2011re, Blas:2011rf}.
We consider, in this work, two combinations of data corresponding to full-shape analyses from large-scale structure data (\textit{i.e.}, the LSS dataset) and from CMB primary data (\textit{i.e.}, the CMB dataset). The full details of these two datasets are provided here:
\begin{itemize}
    \item \textbf{LSS}: This first dataset includes the (one-loop) full-shape analysis of galaxy power spectra in redshift space from two SDSS samples, together with BBN measurements. Here we give the details of the data that compose this dataset:
    \begin{itemize}
        \item \textbf{BOSS DR12 LRG:} The monopole and quadrupole of the galaxy power spectra from BOSS DR12 luminous red galaxies (LRG), cross-correlated with the reconstructed BAO parameters \cite{BOSS:2015fqm}. The SDSS-III BOSS DR12 galaxy sample data and covariances are described in \cite{BOSS:2016wmc,Kitaura:2015uqa}. The measurements, obtained in \cite{Zhang:2021yna}, are from BOSS catalogs DR12 (v5)~\cite{BOSS:2015ewx}. They are divided into four sky cuts, made up of two redshift bins, namely LOWZ with $0.2<z<0.43 \  (z_{\rm eff}=0.32)$, and CMASS with $0.43<z<0.7  \ (z_{\rm eff}=0.57)$, with north and south galactic skies for each, respectively, denoted NGC and SGC.  In the following, we refer to this data simply as ``BOSS''.
        \item \textbf{eBOSS DR16 QSO:} 
        The monopole and quadrupole of the bias tracer power spectra from eBOSS DR16 quasi-stellar objects (QSO) \cite{eBOSS:2020yzd}. The QSO catalogs are described in \cite{eBOSS:2020mzp} and the covariances are built from the EZ-mocks described in \cite{Chuang:2014vfa}. There are about 343 708 quasars selected in the redshift range $0.8<z<2.2$, with $z_{\rm eff}=1.52$, divided into two skies, NGC and SGC~\cite{Beutler:2021eqq,eBOSS:2020gbb}. In the following, we refer to this data simply as ``eBOSS''.
        \item \textbf{BBN:} When we consider only large-scale structure data, we also use the CMB-independent BBN measurement of $\omega_b$ \cite{Schoneberg:2019wmt} that uses the theoretical prediction of \cite{Consiglio:2017pot}, the experimental deuterium fraction of \cite{Cooke:2017cwo}, and the experimental helium fraction of \cite{Aver:2015iza}.
    \end{itemize}
     
    \item \textbf{CMB}: This second dataset includes the low-$\ell$ CMB temperature and polarization auto-correlations (TT, EE), and the high-$\ell$ TT, TE, EE data~\cite{Planck:2019nip} from {\it Planck}~2018~\cite{Planck:2018lbu}. Note that in this analysis, we only consider the CMB primary power spectra, and not the gravitational lensing potential reconstructed from \textit{Planck} (since some of the models considered in this work have been reconstructed without lensing). We have verified that the lensing power spectrum does not affect our conclusions.

\end{itemize}
In this work, we are considering either the LSS dataset, the CMB dataset, or the combination of these two datasets. For the latter, we do not consider the BBN likelihood, since the physical baryon abundance $\omega_b$ is well constrained by \textit{Planck} data \cite{Chu:2004qx,Motloch:2020lhu}.
For all runs performed,\footnote{Except for the regMRL model presented in \cref{sec:regMRL} where $A_s$ and $n_s$ are not present.} we impose large flat priors on the $\Lambda$CDM cosmological parameters $\{ \omega_b,\omega_{\rm cdm}, h, \tau,  \ln 10^{10}A_s, n_s \}$.
For the neutrino treatment, we consider two massless and one massive species with $m_\nu = 0.06 e$V, corresponding to the \planck~convention~\cite{Planck:2018lbu}.
We consider that our chains have converged when the Gelman-Rubin criterion $R-1<0.05$.
Finally, we acknowledge the use of \texttt{Getdist}\footnote{\url{https://getdist.readthedocs.io/en/latest/}} \cite{Lewis:2019xzd} to extract the probability density functions and produce our plots.

\subsection{EFT likelihood}

While the full shape of the CMB primary power spectra are accurately computed within the linear perturbation theory thanks to the Boltzmann solver \classy, we need to go beyond the linear regime to compute the full shape of the (e)BOSS galaxy power spectra. To do so, we use the effective field theory of large-scale structure (EFTofLSS), a semi-numerical method allowing us to compute the galaxy power spectrum in redshift space up to one-loop.\footnote{The first formulation of the EFTofLSS was carried out in Eulerian space in Refs.~\cite{Carrasco:2012cv,Baumann:2010tm} and in Lagrangian space in \cite{Porto:2013qua}. Once this theoretical framework was established, many efforts were made to improve this theory and make it predictive, such as the understanding of renormalization \cite{Pajer:2013jj, Abolhasani:2015mra}, the IR-resummation of the long displacement fields \cite{Senatore:2014vja, Baldauf:2015xfa, Senatore:2014via, Senatore:2017pbn, Lewandowski:2018ywf, Blas:2016sfa}, and the computation of the two-loop matter power spectrum \cite{Carrasco:2013sva, Carrasco:2013mua}. Then, this theory was developed in the framework of biased tracers (such as galaxies and quasars) in Refs. \cite{Senatore:2014eva, Mirbabayi:2014zca, Angulo:2015eqa, Fujita:2016dne, Perko:2016puo, Nadler:2017qto}.}
We use the \pybird\ code \github{https://github.com/pierrexyz/pybird}\cite{DAmico:2020kxu} for the theoretical prediction of this observable as well as for the likelihood of the full-modeling information of (e)BOSS.

\paragraph{EFT priors:}
The EFT one-loop galaxy power spectrum in redshift space possesses 10 free parameters, namely 4 galaxy bias parameters ($b_i$, with $i=[1,4]$),\footnote{Note that \pybird~uses a linear combination of $b_2$ and $b_4$, namely $c_2 = (b_2+b_4)/\sqrt{2}$ and $c_4 = (b_2-b_4)/\sqrt{2}$.} 3 counterterms ($c_\text{ct}$, which is a linear combination of the dark matter sound speed~\cite{Baumann:2010tm,Carrasco:2012cv} and a higher-derivative bias~\cite{Senatore:2014eva}, as well as $c_{r,1}$ and $c_{r,2}$ corresponding to the redshift-space counterterms~\cite{Senatore:2014vja}), and 3 stochastic parameters ($c_{\epsilon,0}$, which is a constant shot noise parameter, as well as $c_{\epsilon}^{\textrm{mono}}$ and $c_{\epsilon}^{\textrm{quad}}$ corresponding to the scale-dependant stochastic contributions of the monopole and the quadrupole).
However, in this study, we set to zero the parameter $c_{r,2}$ (degenerated with $c_{r,1}$, as we do not include the hexadecapole) \cite{DAmico:2019fhj}. In addition, we do not consider $c_4 = (b_2-b_4)/\sqrt{2}$ and $c_{\epsilon}^{\textrm{mono}}$ since the functions that are multiplied by these parameters are negligible compared to the signal-to-noise ratio associated with the (e)BOSS volume (see Ref.~\cite{DAmico:2019fhj}). We finally have 7 EFT parameters per sky cut, implying that we have $7 \times 6 = 42$ EFT parameters for the BOSS and eBOSS data.
Note that in this analysis, we use the so-called ``WC priors'' for the EFT parameters (see Refs.~\cite{DAmico:2019fhj,Colas:2019ret,DAmico:2020kxu}), together with the standard \pybird\ treatment \cite{DAmico:2020kxu}: the EFT parameters that enter linearly in the theory are analytically marginalized within the Gaussian prior $\mathcal{N}(0,2)$, while for the other parameters, namely the ones that do not enter linearly in the theory (\textit{i.e.}, $b_1$ and $c_2 = (b_2+b_4)/\sqrt{2}$), we follow the prescription of Ref.~\cite{DAmico:2022osl}, where we vary $b_1$ within a flat prior $b_1 \in [0,4]$ and $c_2$ within the Gaussian prior $\mathcal{N}(0,2)$.
Given the high-dimensional EFT phase-space, we note that an EFTofLSS analysis applied to the (e)BOSS data is subject to prior volume projection effects (see \textit{e.g.}, Refs. \cite{Simon:2022lde,Carrilho:2022mon,Holm:2023laa,Maus:2023rtr,Gsponer:2023wpm}), which can bias cosmological results. Following Ref.~\cite{Simon:2022lde}, these prior effects can be quantified, for a given parameter $X$, by the following metric:
\begin{equation}
    n\sigma=\frac{X_{\rm mean} - X_{\rm bestfit}}{\sigma_X} \, ,
    \label{eq:projection_effect}
\end{equation}
corresponding to the distance of the mean value $X_{\rm mean}$ from the bestfit value $X_{\rm bestfit}$ (or from the truth of a mock data analysis if we perform such an analysis, as in the following) normalized by the $68\%$ C.L. error bar (of the posterior distribution) $\sigma_X$. 

\paragraph{Scale cut:} In this paper, we analyze the BOSS data up to $k_{\rm max}^{\rm CMASS} = 0.23 h \, {\rm Mpc}^{-1}$ for the CMASS sky cut and up to $k_{\rm max}^{\rm LOWZ} = 0.20 h\, {\rm Mpc}^{-1}$ for the LOWZ sky cut (as determined in Ref.~\cite{Colas:2019ret}), while
we analyze the eBOSS data up to $k_{\rm max}^{\rm eBOSS} = 0.24 h\,{\rm Mpc}^{-1}$ (as determined in Ref.~\cite{Simon:2022csv}). These scales correspond to the maximum wavenumbers for which the one-loop prediction is sufficiently accurate (\textit{i.e.}, above $k_{\rm max}$, we need to take into account the two-loop contribution).
The minimum wavenumber included in our analysis is $k_{\rm min} = 0.01 h \, {\rm Mpc}^{-1}$ for all the samples.

\paragraph{EFT scales:} Regarding the EFT scales, we fix the renormalization scale controlling the spatial derivative expansion (which corresponds to the typical extension of the host halo) to $k_\textsc{m} = k_\textsc{nl} = 0.7 \, h {\rm Mpc}^{-1} $ for BOSS and eBOSS. In addition, we fix the renormalization scale controlling the velocity expansion (appearing in the redshift-space expansion) to $k_R = 0.35 \, h {\rm Mpc}^{-1}$ for BOSS and to  $k_R =  0.25 \, h {\rm Mpc}^{-1}$ for eBOSS.
Finally, we fix the mean galaxy number density to $\Bar{n}_g = 4 \cdot 10^{-4} \, ({\rm Mpc}/h)^3$ ($\Bar{n}_g = 2 \cdot 10^{-5} \, ({\rm Mpc}/h)^3$) for BOSS (eBOSS).
These scales have been determined in Refs.~\cite{DAmico:2019fhj,DAmico:2021ymi} for BOSS and in Ref.~\cite{Simon:2022csv} for eBOSS.

\paragraph{Observational effects:} Finally, our analysis also takes into account several observational effects (see Ref.~\cite{DAmico:2019fhj}), such as the Alcock-Paczyński effect~\cite{Alcock:1979mp}, the window functions as implemented in Ref.~\cite{Beutler:2018vpe} (see App.~A of Ref.~\cite{Simon:2022adh} for more details), and binning~\cite{DAmico:2022osl}.

\section{The standard model: power law primordial spectrum}\label{sec:powerlaw}
\begin{figure}
    \centering
    \includegraphics[width=0.49\textwidth]{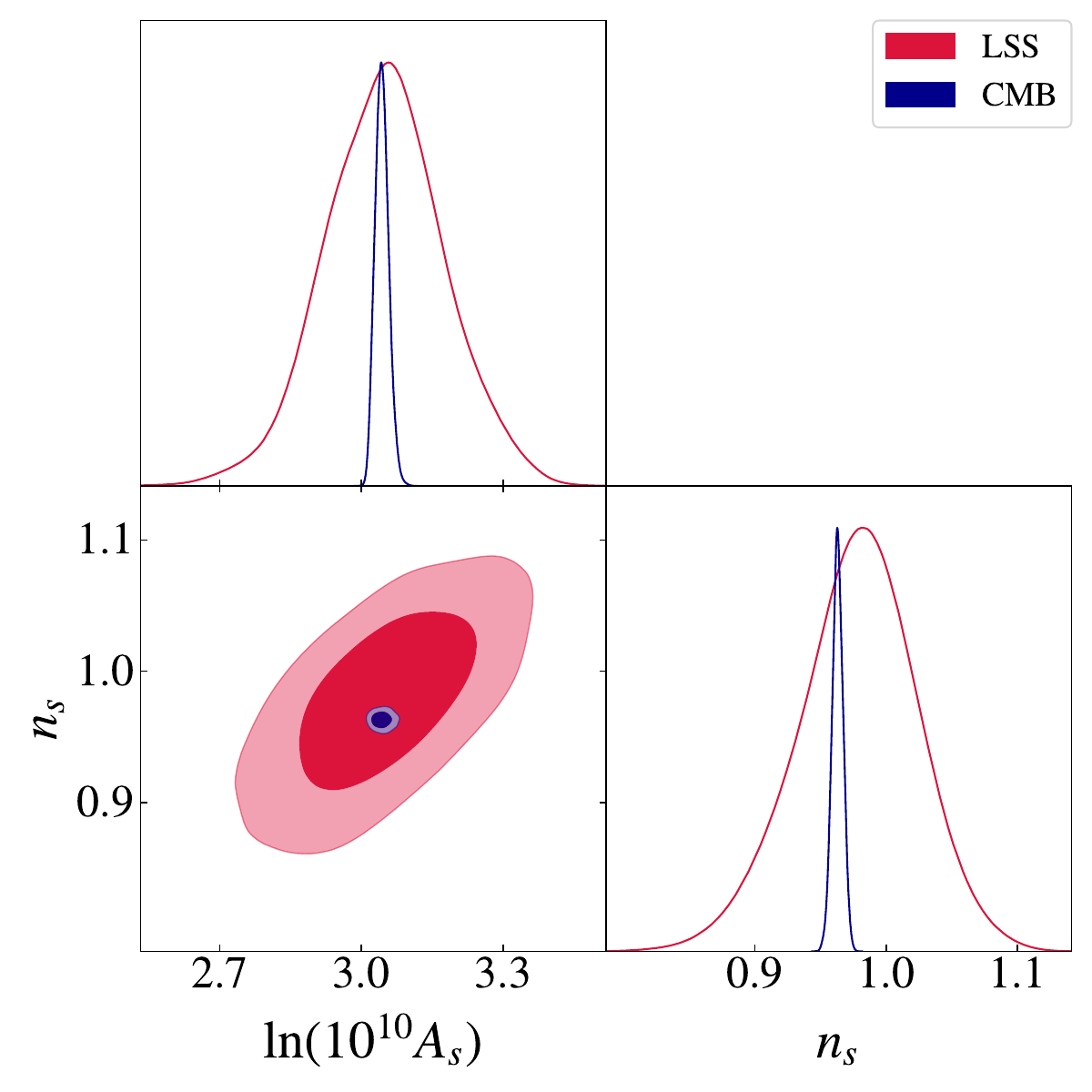}
    \includegraphics[width=0.49\textwidth]{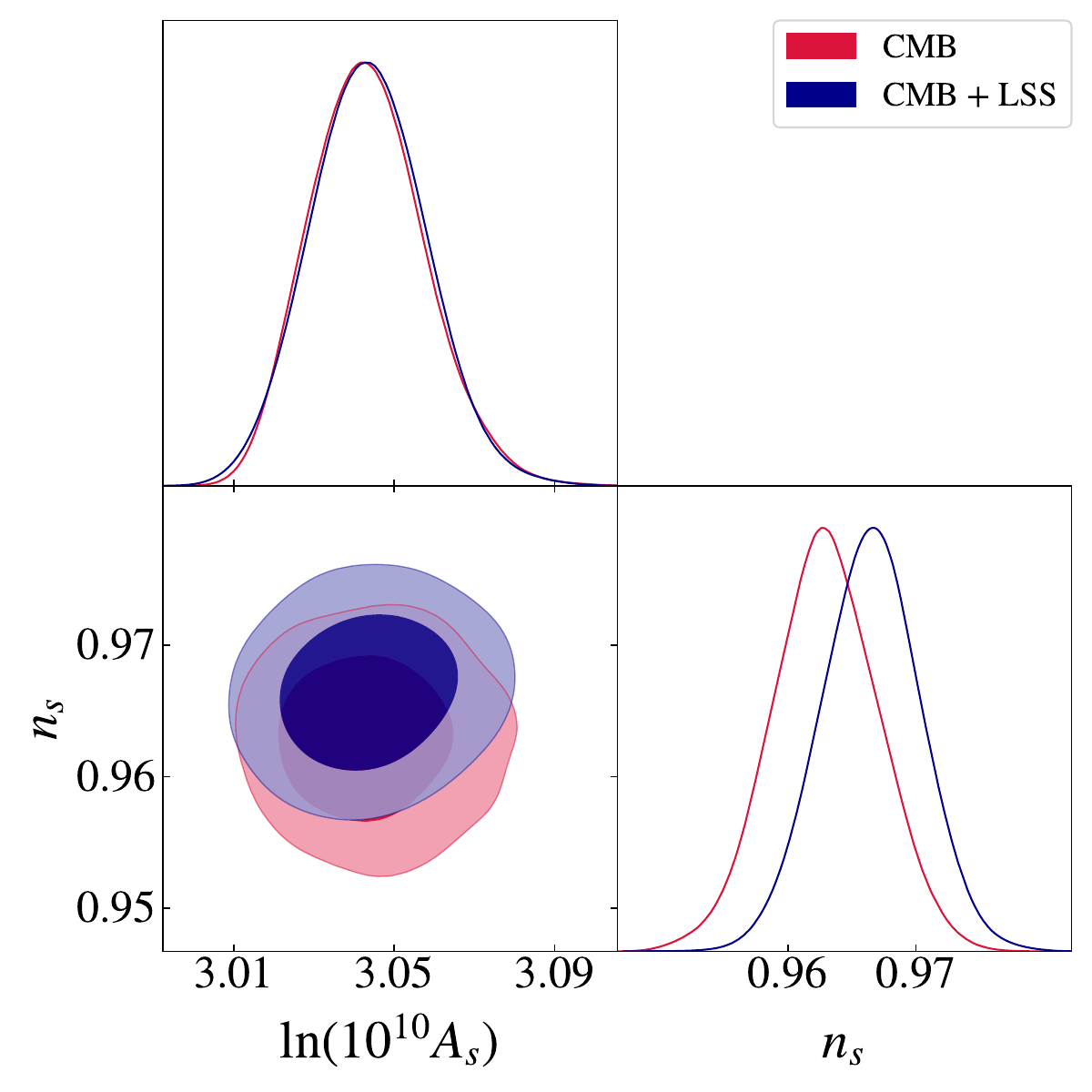}
    \caption{\textit{Left:} 1D and 2D posterior distributions reconstructed from the LSS and CMB datasets for the standard power-law parametrization. \textit{Right:} Same with the CMB and CMB + LSS datasets.}
    \label{fig:power_law}
\end{figure}

We begin by presenting our results for the standard $\Lambda$CDM model, assuming a simple power law form of the primordial spectrum,
\begin{equation}\label{eq:PL}
    \mathcal{P}_{\zeta,0}(k)=A_s\left(\frac{k}{k_*}\right)^{n_s-1}~,
\end{equation}
where $k_*=0.05 ~\rm Mpc^{-1}$ is the pivot scale, and where $A_s$ and $n_s$ are the amplitude and the tilt of the primordial power spectrum, respectively.

In \cref{fig:power_law}, we present the posterior distributions of $[\ln (10^{10}A_s), n_s]$ from our LSS, CMB and CMB + LSS datasets.
In particular, for the LSS dataset, we find $\ln (10^{10}A_s) = 3.05\pm 0.13$ and $n_s = 0.978^{+0.048}_{-0.041}$, to be compared with the \textit{Planck} constraints, namely $\ln (10^{10}A_s) = 3.045\pm  0.016 $ and $n_s = 0.9649 \pm 0.0044$ (corresponding to the ``TT,TE,EE+lowE'' results of Ref.~\cite{Planck:2018lbu}).
Although the \textit{Planck} constraints are respectively $\sim 8$ and $\sim 10$ times better than those of the LSS dataset, it is interesting to note that the LSS constraints alone are completely consistent with the \textit{Planck} CMB measurements.
In addition, given that the CMB and LSS results are compatible, we can combine these two datasets (as done in the right panel of \cref{fig:power_law}), and show that the addition of the LSS dataset on top of the CMB dataset is able to slightly improve the constraint on $n_s$ by $7 \%$, while the constraint on $\ln (10^{10}A_s)$ is not affected.

This standard analysis allows us to
emphasize two important points: (i) LSS data are capable of providing information on the initial conditions of the Universe, and (ii) this information does not improve our understanding of the initial conditions compared with the CMB.
In the remainder of this paper, we will therefore investigate whether an EFT analysis of the (e)BOSS data can be used to constrain non-standard behaviour of the PPS and provide additional information compared with the CMB.
The main goal of this paper is twofold: (i) to gauge the sensitivity of the current large-scale structure data on inflationary dynamics beyond the standard ``slow-roll'' regime,
and (ii) to assess how much additional information we can gain compared with CMB analysis alone.

\section{Global features: linear and logarithmic oscillations}\label{sec:Global}

Many physically-motivated models of inflation\textemdash as well as many  alternative early universe scenarios \cite{Chen_2011,Chen:2015lza,Chen:2018cgg,Quintin:2024boj}\textemdash naturally give rise to oscillations, or ``features'', on top of the simple power law described by \cref{eq:PL}.
In particular, sudden changes in the background quantities during the inflationary epoch lead to so-called ``global features'', spanning a wide range of scales. 
Among other,  examples leading to such features include a discontinuity or ``step'' in the inflaton potential \cite{Starobinsky:1992ts,Adams:2001vc,Gong:2005jr,Hazra:2010ve,Hazra:2014goa,Petretti:2024mjy} or a non-standard choice of vacuum \cite{Danielsson:2002kx,Easther:2002xe,Martin:2003kp}. Logarithmic oscillations, on the other hand, usually arise when there is an oscillatory pattern in the potential itself, as for periodic potentials, which are ubiquitous in high-energy physics.
One interesting and commonly studied inflationary model leading to global logarithmic oscillations in the PPS is the \textit{axion monodromy} \cite{Chen:Osc000,Silverstein:2008sg,Flauger:2009ab,Aich:Osc3,Kaloper:2011jz,Flauger:2014ana,Peiris:2013opa}. 

Beyond theoretical motivations, the study of linear and logarithmic oscillations superimposed on a power-law spectrum offers valuable pedagogical insights. These oscillatory features have been extensively explored in the literature due to their simplicity and theoretical relevance (see \textit{e.g.}, Refs.~\cite{Planck:2015sxf,Planck:2018jri} in the context of CMB analyses, and Refs.~\cite{Beutler:2019ojk,Ballardini:2019tuc,Ballardini:2022wzu,Mergulhao:2023ukp} in the context of LSS analyses). 
Here, following Ref.~\cite{Chen:2008wn} and to remain model agnostic, we perform the usual (template-based) search for features \cite{Planck:2015sxf,Planck:2018jri}
\begin{equation}\label{eq:log-osc}
\mathcal{P}_\zeta(k)=\mathcal{P}_{\zeta,0}(k)\left[1+A_{X}\sin\left(\omega_{X}~\mathcal{K}_X+\phi_{X}\right)\right]~,
\end{equation}
where  $\mathcal{P}_{\zeta,0}$ is the standard power-law given by \cref{eq:PL}, $X=\{\rm lin, log\}$, and where $A_{X}$, $\omega_{X}$, and $\phi_{X}$ are respectively the amplitude, frequency, and phase of the oscillations. For linear oscillations, we have $\mathcal{K}_{\rm lin}=k/k_*$, while for logarithmic oscillations, $\mathcal{K}_{\rm log}=\ln\left(\frac{k}{k_*}\right)$. We note that in order to resolve the log-oscillations in the matter power spectrum properly, we have worked with increased precision settings $\texttt{k\_per\_decade\_for\_pk}=\texttt{k\_per\_decade\_for\_bao}=200$ in the Boltzmann solver \classy.

\subsection{The priors}\label{sec:priors}

\begin{table}[t]
    \centering
    \begin{tabular}{lccc}
    \hline\hline
     Model & Parameter & Prior LSS & Prior CMB\\  
     \hline
     & $A_\mathrm{lin}$ &  $\mathcal{U}[0, 0.5]$  &  $\mathcal{U}[0, 0.5]$\\
   Linear oscillations & $\log_{10}\omega_\mathrm{lin}$ &  $\mathcal{U}[0.3, 1.35]$ &  $\mathcal{U}[0, 2.0]$  \\
    & $\phi_\mathrm{lin}$ &  $\mathcal{U}[0, 2\pi]$ &  $\mathcal{U}[0, 2\pi]$  \\
    \hline
     & $A_\mathrm{log}$ &  $\mathcal{U}[0, 0.5]$  &  $\mathcal{U}[0, 0.5]$\\
    Logarithmic oscillations& $\log_{10}\omega_\mathrm{log}$ &  $\mathcal{U}[0.7, 1.9]$  &  $\mathcal{U}[0, 2.1]$\\
    & $\phi_\mathrm{log}$ &  $\mathcal{U}[0, 2\pi]$  &  $\mathcal{U}[0, 2\pi]$\\
    \hline\hline
    \end{tabular}
    \caption{Priors used in the linear and logarithmic oscillation analyses. Let us note that for the CMB + LSS analysis we use the CMB priors of Ref.~\cite{Planck:2018jri}.}
    \label{tab:priors_lin_log}
\end{table}

Before presenting the results, let us start by discussing the priors, displayed in \cref{tab:priors_lin_log}, used in the linear and logarithmic oscillation analyses. The priors considered for the CMB dataset match that of the \planck~analysis \cite{Planck:2018jri}. For the LSS dataset, we can apply these priors on the amplitude $A_{\rm X}$ and the phase $\phi_\mathrm{\rm X}$, but not on the frequency $\omega_{\rm X}$.
Indeed, the frequency range that our LSS analysis can probe is restricted by two limiting factors.
To understand this, let us start with the linear oscillation scenario.
First, our analysis is not sensitive to oscillations whose period exceeds the maximum wavenumber used in our analysis. There is therefore a lower limit of the frequencies to which we are sensitive due to the total $k$-range ($\Delta k^{\rm tot} = k_{\rm max} - k_{\rm min} \sim k_{\rm max}$) of the BOSS and eBOSS data that we can analyze with the one-loop EFTofLSS.
In particular, the minimum frequency we can probe is given by $\omega_{\rm lin}^{\rm min} = 2 \pi k_* / \Delta k ^{\rm tot}$. Considering that the largest $k$-range comes from eBOSS, namely $\Delta k^{\rm tot}_{\rm eBOSS} = 0.24 h{\rm Mpc}^{-1}$, and that $h = 0.6736$ (from Ref.~\cite{Planck:2018vyg}), we obtain $\omega_{\rm lin}^{\rm min} = 1.9$. In order to remain conservative, we choose $\omega_{\rm lin}^{\rm min} = 2$ (or $\log_{10}\omega_{\rm lin}^{\rm min} = 0.3$).
Second, our analysis is not sensitive to oscillations with a period which is too small compared to the distance $\Delta k$ between the data points. There is therefore an upper limit of the frequencies to which we are sensitive due to the finite survey volume of galaxy clustering experiments.
The maximum frequency is given by the Nyquist-Shannon theorem \cite{1697831}, which states that the signal can only be determined if its frequency is less than half the frequency of the data, namely $\omega_{\rm lin}^{\rm max} = \pi k_* / \Delta k$. Given that in our analysis $\Delta k = 0.01 h{\rm Mpc}^{-1}$, we obtain $\omega_{\rm lin}^{\rm max} = 23.3$. In order to remain conservative, we choose $\omega_{\rm lin}^{\rm max} = 22.5$ (or $\log_{10}\omega_{\rm lin}^{\rm max} = 1.35$).

\begin{figure}
    \centering
    \includegraphics[width=\textwidth]{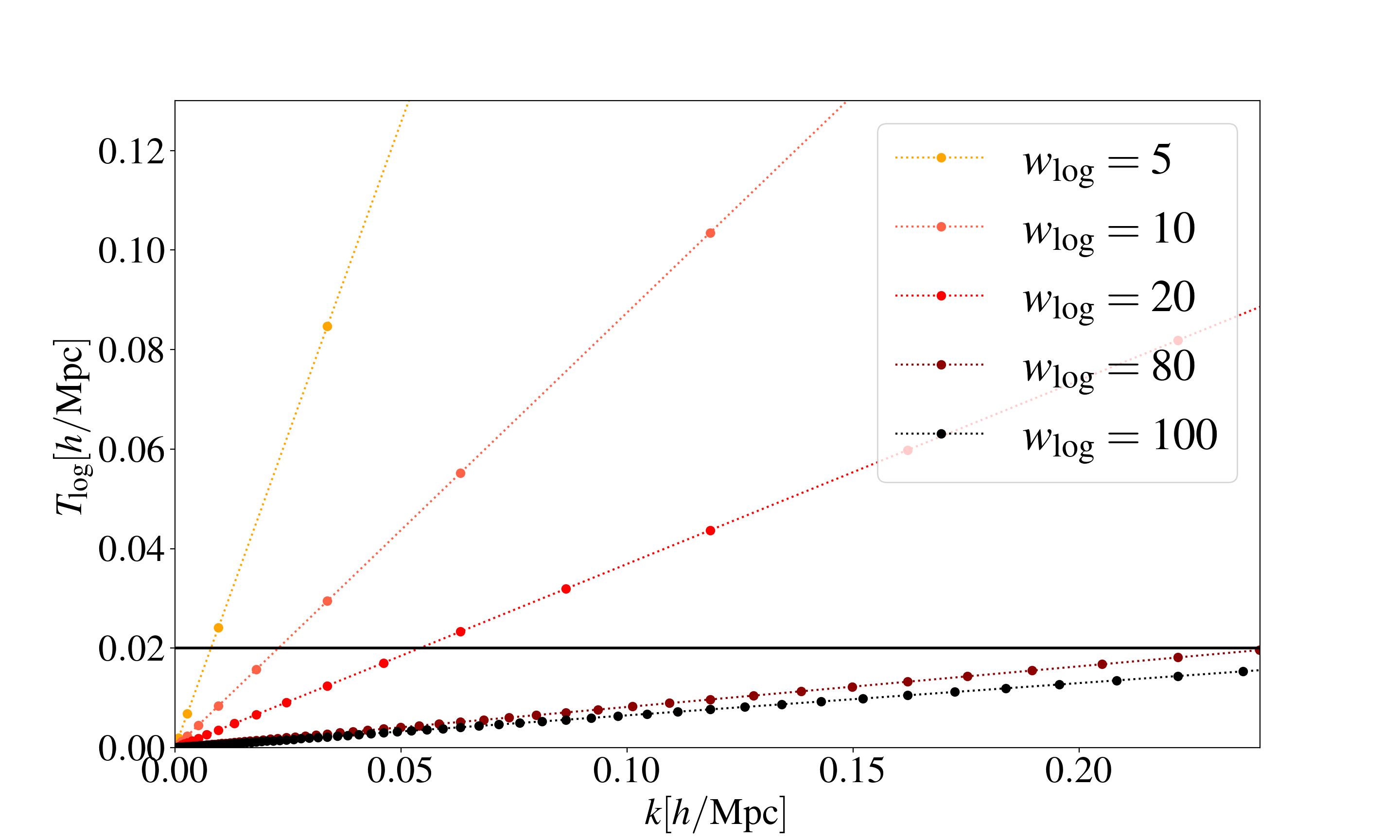}
    \caption{Logarithmic oscillation period $T_{\rm log}$ as a function of the wavenumber $k$ for different values of $\omega_{\rm log}$. In particular, the distance between two points corresponds to one period $T_{\rm log} = k_* \left( e^{2 \pi (n+1)/ \omega_{\rm log}} - e^{2 \pi n/ \omega_{\rm log}} \right)$. The black horizontal line represents the Nyquist-Shannon period $T_{\rm log}^{\rm NS} = 2 \times \Delta k = 0.02 h/{\rm Mpc}$.}
    \label{fig:log_priors}
\end{figure}

For the logarithmic scenario, the determination of $\omega_{\rm log}^{\rm min}$ and $\omega_{\rm log}^{\rm max}$ is slightly more involved, as the frequency of the signal depends on the wavenumber $k$. Following Ref.~\cite{Beutler:2019ojk}, we represent in \cref{fig:log_priors} the logarithmic oscillation period $T_{\rm log}$ as a function of the wavenumber $k$ for different values of $\omega_{\rm log}$. 
To determine $\omega_{\rm log}^{\rm min}$, we seek the minimum value of the frequency that allows us to have a complete period in $\Delta k^{\rm tot}_{\rm eBOSS} = 0.24 h{\rm Mpc}^{-1}$. This determines the minimum frequency to be $\omega_{\rm log} = 3$, but we choose $\omega_{\rm log}^{\rm min} = 5$ ($\log_{10} \omega_{\rm log}^{\rm min} = 0.7$) to remain conservative.
To determine $\omega_{\rm log}^{\rm max}$, we need to find the maximum frequency that allows us to have a complete period above the Nyquist-Shannon period (shown as a black horizontal line in \cref{fig:log_priors}). This is achieved for $\omega_{\rm log} < \omega_{\rm log}^{\rm max} = 80$ ($\log_{10} \omega_{\rm log}^{\rm max} = 1.9$). In \cref{fig:lin_prior_lin_log} of \cref{app:lin_log}, we show that the constraints on $A_{\rm X}$ do not depend on whether the prior on $\omega_{\rm X}$ is linear or logarithmic. \\

Let us note that we can improve the frequency range in our analysis by decreasing the value of the bandwidth $\Delta k$, as performed in Refs.~\cite{Beutler:2019ojk,Mergulhao:2023ukp}, which would allow us to include higher frequencies.
In particular, the authors of Ref.~\cite{Beutler:2019ojk} use a bandwidth of $\Delta k = 0.005 h{\rm Mpc}^{-1}$, allowing them to consider $\omega_{\rm lin}^{\rm max} =45$ and $\omega_{\rm log}^{\rm max} = 80$, while the authors of Ref.~\cite{Mergulhao:2023ukp} use a bandwidth of $\Delta k = 0.001 h{\rm Mpc}^{-1}$, allowing them to consider $\omega_{\rm lin}^{\rm max} = 200$ and $\omega_{\rm log}^{\rm max} = 360$.
We leave for future work the application of our EFTofLSS analysis to (e)BOSS data with a smaller bandwidth.
However, we would like to emphasize that our analysis can probe smaller frequencies than Refs.~\cite{Beutler:2019ojk,Mergulhao:2023ukp}, which makes it complementary.
This is due to the fact that, unlike Refs.~\cite{Beutler:2019ojk,Mergulhao:2023ukp}, we take into account the effects (on the full-shape) of the small-scale non-linearities. 
Our minimum frequency is therefore determined by the maximum wavenumber that  can be analyzed with the EFTofLSS, while in Refs.~\cite{Beutler:2019ojk,Mergulhao:2023ukp} the minimum frequency is determined by the wavenumber at which small-scale non-linearities become non-negligible (at $k \sim 0.15 h/{\rm Mpc}$), implying that they consider $\omega_{\rm lin}^{\rm min} =5$ and $\omega_{\rm log}^{\rm min} = 10$.

\subsection{Validation with N-body and IR-resummation} \label{sec:Nbody}

\begin{figure}
    \centering
    \includegraphics[width=\linewidth]{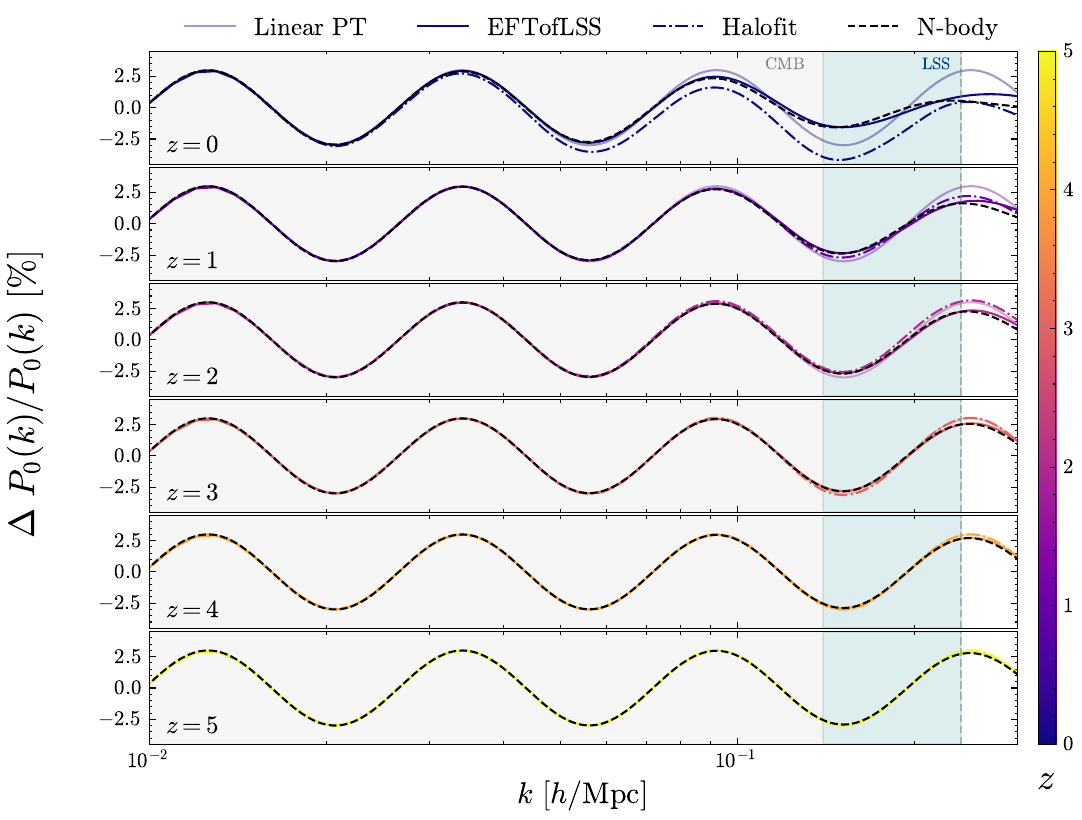}
    \caption{Non-linear matter power spectrum residuals of the logarithmic oscillations (see \cref{eq:log-osc}) with respect to a featureless power-law, considering an amplitude $A_{\rm log}=0.03$, a frequency $\log_{10}\omega_{\rm log}=0.8$, and a phase $\phi_{\rm log}=0.2$.
    We show the prediction from the linear perturbation theory, the EFTofLSS, \texttt{Halofit}, as well as from the analytical fitting formulae of \cite{Ballardini:2019tuc}, based on N-body simulations.
    This figure shows an excellent agreement between the EFTofLSS prediction and the N-body simulations. \cref{fig:Nbody-comp-lin} shows the equivalent for the case of linear oscillations, where we also find a good agreement with the N-body results.}
    \label{fig:Nbody-comp}
\end{figure}

Before fitting real data, we start by comparing the one-loop EFT prediction for the matter power spectrum against the results from N-body simulations of Ref.~\cite{Ballardini:2019tuc} for both linear and logarithmic (global) features, providing a valuable consistency check and validating our analysis pipeline.\footnote{Note that the accuracy of the EFTofLSS in capturing non-linear effects, including two-loop corrections, was demonstrated in \cite{Foreman:2015lca} for the matter power spectrum. In this work, we extend that validation by explicitly showing that the EFTofLSS (at one-loop) remains reliable for the scales probed by our dataset, even in the presence of non-standard initial conditions that introduce oscillatory features in the power spectrum.}
In \cref{fig:Nbody-comp}, we show the matter power spectrum residuals (with respect to $\Lambda$CDM) as a function of redshift for logarithmic oscillations  with an amplitude $A_{\rm log}=0.03$, a frequency $\log_{10}\omega_{\rm log}=0.8$, and a phase $\phi_{\rm log}=0.2$ (see \cref{eq:log-osc}). These values have been chosen to match those of Ref.~\cite{Ballardini:2019tuc}, allowing for easier comparison between our results and theirs.
In \cref{fig:Nbody-comp-lin} of \cref{app:lin_log}, we produce the same figure for linear oscillations.
We show that the one-loop EFTofLSS prediction (accounting for IR-resummation) from \pybird\ is in excellent agreement at $<0.5\%$ with the semi-analytical fitting formulae based on N-body simulations (dashed black lines) from Ref.~\cite{Ballardini:2019tuc}. 
We confirm the findings from previous literature \cite{Vlah:2015zda,Vasudevan:2019ewf,Beutler:2019ojk,Mergulhao:2023ukp} suggesting that non-linearities introduce a damping of the oscillatory features, as shown in \cref{fig:Nbody-comp,fig:Nbody-comp-lin} when comparing the linear prediction with the EFTofLSS prediction.
We note that this damping is clearly redshift-dependent, since it is enhanced for small redshifts (\textit{i.e.}, when the non-linearities become more important).

Additionally, our results confirm that the \texttt{Halofit} prescription struggles to accurately reproduce N-body results in the presence of primordial features, particularly at low redshift and on small scales. Given the data precision, this discrepancy could introduce non-negligible shifts in the posterior distributions when analyzing models with primordial features. Certain precautions must therefore be taken when analyzing the CMB lensing with \texttt{Halofit} to model non-linear corrections.

\begin{figure}[ht]
    \centering
    \includegraphics[width=\linewidth]{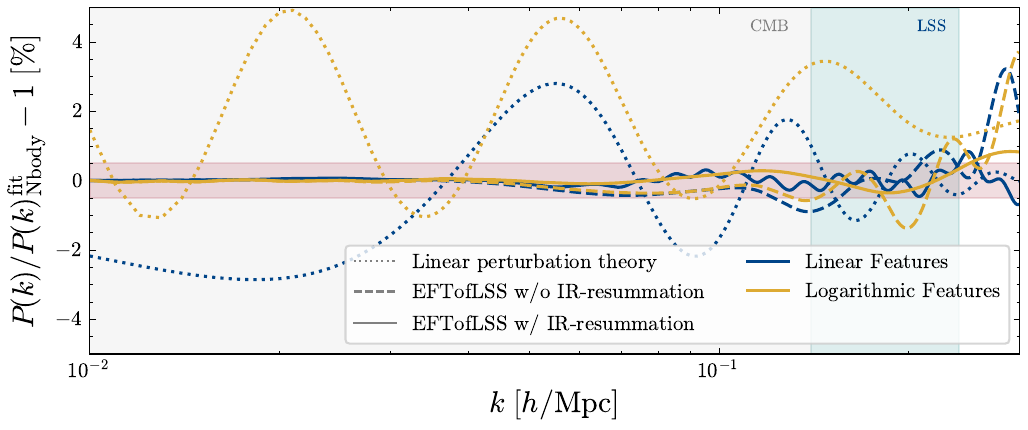}
    \caption{Matter power spectrum residuals at $z=0$, relative to the N-body prediction of Ref.~\cite{Ballardini:2019tuc}, for the linear and logarithmic oscillations. We compare the non-linear (one-loop) EFT predictions with (solid lines) and without (dashed lines) accounting for the IR-resummation, as implemented in \pybird. When accounting for the long-wavelength displacements, the one-loop prediction (solid lines) shows an excellent agreement (at $<0.5\%$) with the results from N-body simulations~\cite{Ballardini:2019tuc} up to $k_{\rm max}=0.23~h{\rm Mpc}^{-1}$\textemdash as indicated by the red-shaded region\textemdash while maintaining sub-percent accuracy up to $k_{\rm max}=0.3~h{\rm Mpc}^{-1}$. 
    }
    \label{fig:comparison_Nbody_IR}
\end{figure}

To further investigate the excellent agreement between our results and the N-body simulations, we plot in \cref{fig:comparison_Nbody_IR} the residuals between our prediction from \pybird~and the fitting formula from Ref.~\cite{Ballardini:2019tuc}.
In particular, we isolate in our prediction the non-linear effects originated by the UV physics (through the EFTofLSS contributions) from those originated by the long-wavelength displacements (through the IR-resummation). While the former are integrated out and treated perturbatively thanks to the one-loop EFTofLSS computation, the latter are non-perturbative and therefore need to be resummed to accurately describe the oscillatory features in the power spectrum~\cite{Senatore:2014via} (including the BAO wiggles and the primordial oscillations).

First, we can see in \cref{fig:comparison_Nbody_IR} that the EFTofLSS corrections from UV physics allow us to significantly improve the fit to the N-body simulations, since the relative error goes from $<5 \%$ to $< 1.5 \%$ when we include the one-loop EFTofLSS corrections (in dashed line) on top of the linear matter power spectrum (in dotted line). 
Let us recall that the matter power spectrum computed using the EFT at one-loop reads~\cite{Carrasco:2012cv,Baumann:2010tm}
\begin{equation}
    P_{\rm m}(k,a) = P_{11}(k,a) + P_{22}^{\Lambda}(k,a) + 2 \cdot P_{13}^{\Lambda}(k,a) - 2 \cdot c_{\rm s, eff}^2(a) \frac{k^2}{k_{\rm NL}^2} P_{11}(k,a) \ ,
\end{equation}
where $P_{11}$ corresponds to the linear matter power spectrum (\textit{i.e.}, the \textit{tree-level} contribution), and where $P_{22}^{\Lambda}$ and $P_{13}^{\Lambda}$ correspond to the \textit{one-loop} contributions which are integrated over the modes smaller than $\Lambda^{-1}$, corresponding to the EFTofLSS cutoff scale.
The last term corresponds to the counterterm renormalizing $P_{13}^{\Lambda}$,\footnote{Let us note that $P_{22}^{\Lambda}$ is renormalized by a higher-order contribution which is not included in our analysis~\cite{Carrasco:2012cv,Baumann:2010tm}.} and depends on the \textit{effective sound speed} $c_{\rm s, eff}$, which must be adjusted to the data or N-body simulations. The size of this counterterm is controlled by $k_{\rm NL} = 0.7~ h {\rm Mpc}^{-1}$, corresponding to the non-linear scale. 
In \cref{fig:comparison_Nbody_IR}, we have fixed $c_{\rm s, eff}^2 =1$ throughout, which provides a very good agreement with N-body.\footnote{We stress, however, that this is a redshift-dependent quantity; its value should be determined by the data (as done in the MCMC analyses), and, in principle, it can be different for the linear and logarithmic oscillation scenarios.}
Note that \cref{fig:comparison_Nbody_IR} represents the matter power spectrum at $z = 0$, where non-linear effects are most significant, implying a better agreement for the (effective) redshift range considered in this work, namely $z = 0.32 - 1.52 $.

Second, we also display in \cref{fig:comparison_Nbody_IR} the contribution of the IR-resummation (in solid line), and show that we further improve the fit to the N-body simulations, with a relative error of $< 0.5 \%$, shown as the red shaded horizontal region. Given this very good agreement, we validate our pipeline at the level of the matter power spectrum. A careful examination, however, reveals the presence of small residual (high-frequency) oscillations in the EFTofLSS predictions, when compared to the N-body fit of Ref.~\cite{Ballardini:2019tuc}. These tiny discrepancies may arise from the fact that the semi-analytical fitting formula used in Ref.~\cite{Ballardini:2019tuc} assumes a phenomenological Gaussian damping of the oscillations, which may not fully capture finer structures in the power spectrum coming from non-linear processes. Further investigation, including a detailed comparison with N-body simulations, is necessary to determine whether these oscillations are physical and accurately modeled within the EFTofLSS framework. However, given the precision of current and near-future large-scale structure surveys, these $<0.5\%$ residual differences are well within observational uncertainties and do not pose a significant concern for our analysis.

Let us note that unlike the previous analyses~\cite{Beutler:2019ojk,Mergulhao:2023ukp,Ballardini:2022wzu}, our IR-resummation is not based on the wiggle/no-wiggle split procedure, proposed in Refs.~\cite{Blas:2016sfa,Ivanov:2018gjr}, and generalized to inflationary oscillating features in Refs.~\cite{Beutler:2019ojk,Vasudevan:2019ewf,Ballardini:2024dto}, which is used to isolate the BAO part from the broadband shape of the power spectrum.
In this IR-resummation scheme, the power spectrum is separated into two contributions, $P(k)=P^{\rm nw}(k) + P^{\rm w}(k)$, where $P^{\rm nw}(k)$ and $P^{\rm w}(k)$ are respectively the smooth (``no-wiggle'') and the oscillatory (``wiggle'') part of the matter power spectrum. 
In $\Lambda$CDM, the wiggle part encodes the BAO signal, and the IR-resummation correction is only applied to this contribution in order to accurately describe the BAO signal. 
However, if we postulate  non-standard oscillations in the power spectrum, then $P^{\rm w}(k)$ also encodes those features, and we need to modify the IR-resummation accordingly, as done in Refs.~\cite{Beutler:2019ojk,Vasudevan:2019ewf,Ballardini:2024dto} (see also Ref.~\cite{Vlah:2015zda}).
Instead, the \pybird~code utilizes the IR-resummation procedure proposed in Ref.~\cite{Senatore:2014via}, generalized to redshift space in Ref.~\cite{Lewandowski:2015ziq}, and made numerically practical in Ref.~\cite{DAmico:2020kxu}. 
In this approach, the bulk displacements are resummed directly on the full shape (and not only on the wiggly part of the power spectrum). 
This scheme is very advantageous for the study of primordial features since it relies on an analytical procedure that can be applied (in principle) to any shape of the linear matter power spectrum, implying that we do not need to modify the IR-resummation scheme in our analysis.
Ref.~\cite{Ballardini:2019tuc} shows that their N-body simulations are in good agreement with the wiggle/no-wiggle split procedure of  Refs.~\cite{Beutler:2019ojk,Vasudevan:2019ewf} (see also Ref.~\cite{Ballardini:2024dto}), while we explicitly show here that this is also the case for the \pybird~IR-resummation procedure, showing a good consistency between our results and those in past literature.
This corroborates the results of Ref.~\cite{Chen:2020ckc} which found a good agreement between the IR-resummation procedure implemented in \pybird~(based on the Lagrangian perturbation theory) and the wiggle/no-wiggle split procedure (based on the Eulerian perturbation theory).

To be more precise, in \pybird, the IR-resummation of the galaxy power spectrum up to the $N$-loop order $P^\ell(k)_{|N}$ is defined thanks to the $j$-loop order pieces of the Eulerian (\emph{i.e.}, non-resummed) correlation function $\xi^{\ell}_j (k)$ as~\cite{Senatore:2014vja, Lewandowski:2015ziq,DAmico:2020kxu}
\begin{equation} \label{eq:IRresum}
P^\ell(k)_{|N} = \sum_{j=0}^N \sum_{\ell'}  4\pi (-i)^{\ell'} \int dq \, q^ 2 \, Q_{||N-j}^{\ell \ell'}(k,q) \, \xi^{\ell'}_j (q) \ ,
\end{equation}
where $Q_{||N-j}^{\ell \ell'}(k,q)$ captures the effects of the IR-displacements on the galaxy power spectrum:
\begin{align}
&Q_{||N-j}^{\ell \ell'}(k,q) = \frac{2\ell+1}{2} \int_{-1}^{1}d\mu_k \,\frac{i^{\ell'}}{4 \pi} \int d^2 \hat{q} \, e^{-i\vect{q} \cdot \vect{k}} \, F_{||N-j}(\vect{k},\vect{q}) \mathcal{L}_\ell(\mu_k) \mathcal{L}_{\ell'}(\mu_q) \, , \label{eq:resumQ}\\
&F_{||N-j}(\vect{k},\vect{q}) = T_{0,r}(\vect{k},\vect{q})\times T_{0,r}^{-1}{}_{||N-j}(\vect{k},\vect{q}) \, , \nonumber\\
&T_{0,r}(\vect{k},\vect{q}) = \exp \left\lbrace -\frac{k^2}{2} \left[ \Xi_0(q) (1+2 f \mu_k^2 + f^2 \mu_k^2) + \Xi_2(q) \left( (\hat k \cdot \hat q)^2 + 2 f \mu_k \mu_q (\hat k \cdot \hat q) + f^2 \mu_k^2 \mu_q^2 \right) \right] \right\rbrace \,  , \nonumber
\end{align}
with
\begin{align}\label{eq:Xi_02}
\Xi_0(q) & = \frac{2}{3} \int \frac{dp}{2\pi^2} \, \exp \left(-\frac{p^2}{\Lambda_{\rm IR}^2} \right)  P_{11}(p) \, \left[1 - j_0(pq) - j_2(pq) \right] \, , \\
\Xi_2(q) & = 2 \int \frac{dp}{2\pi^2}\, \exp \left( -\frac{p^2}{\Lambda_{\rm IR}^2} \right)  P_{11}(p) \, j_2(pq) \, .
\end{align}
In these equations, $f\equiv \mathrm{d}\ln{D_+(a)}/\mathrm{d\ln{a}}$ is the linear growth rate (where $D_+(a)$ is the growing mode solution of the density contrast), $\mathcal{L}_{\ell}$ is the Legendre polynomial of order $\ell$, and $\mu_k \equiv \hat{z}\cdot \hat{k}$ ($\mu_q \equiv \hat{z}\cdot \hat{q}$) is the angle between the line-of-sight $\vect{z}$ and the wavevector $\vect{k}$ ($\vect{q}$).
In addition, $\Lambda_{\rm IR}$ corresponds to the scale up to which the IR modes are resummed \cite{Lewandowski:2015ziq}. In \pybird, $\Lambda_{\rm IR} = 0.2 h {\rm Mpc}^{-1}$ in order to avoid the effects of some uncontrolled UV modes.
We have verified that the resummed galaxy power spectrum converges smoothly towards our baseline prediction when $\Lambda_{\rm IR} \to 0.2 h {\rm Mpc}^{-1}$, suggesting that we have no numerical instabilities of the IR-resummation scheme for cosmologies where the primordial power spectrum exhibits linear or logarithmic oscillations.

\subsection{Cosmological constraints} \label{sec:global_constraints}

\begin{figure}
    \centering
    \includegraphics[width=0.8\textwidth]{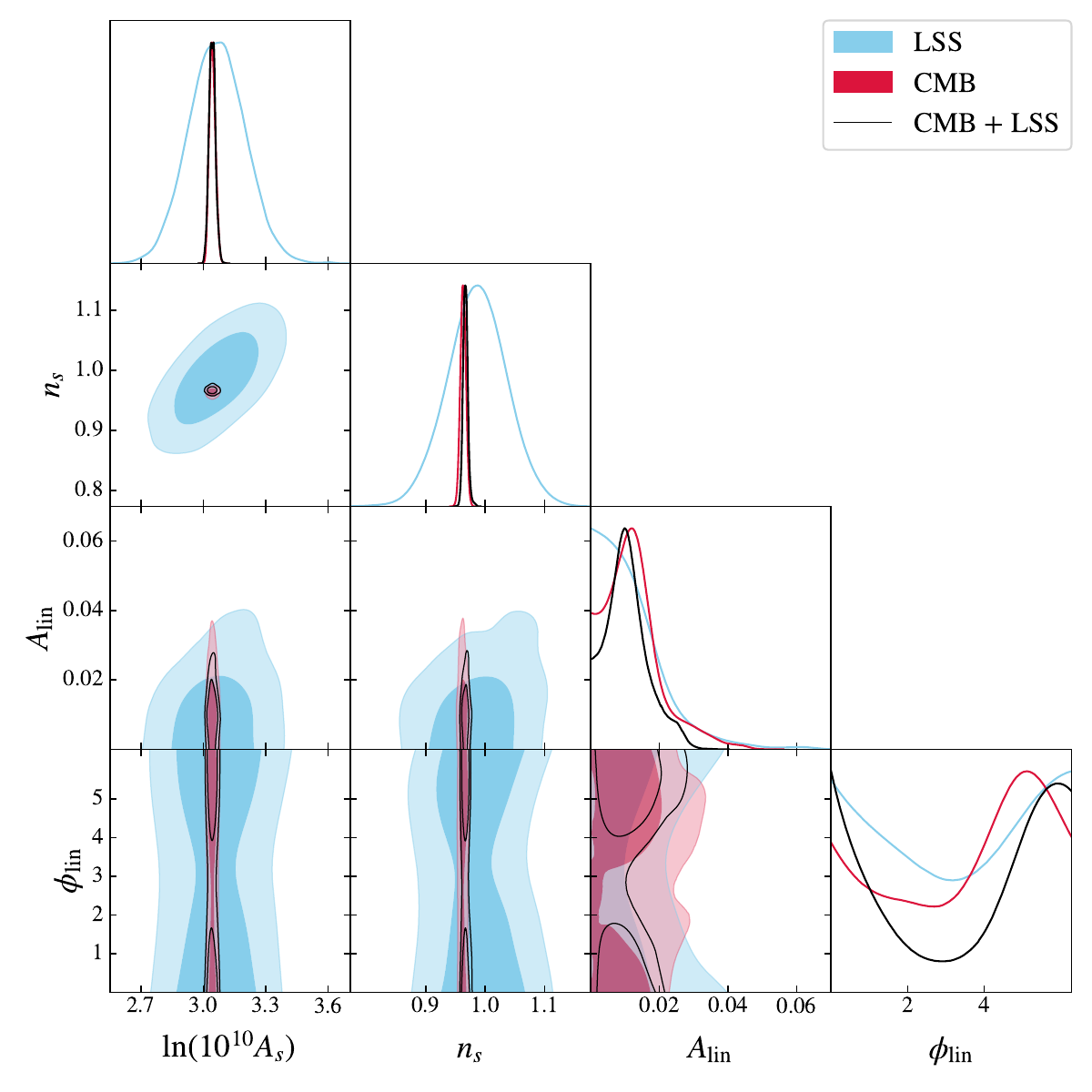}
    \caption{1D and 2D posterior distributions reconstructed from the LSS, CMB, and CMB + LSS datasets for the linear oscillations analysis.}
    \label{fig:lin_oscillations}
\end{figure}

\begin{figure}
    \centering
    \includegraphics[width=0.8\textwidth]{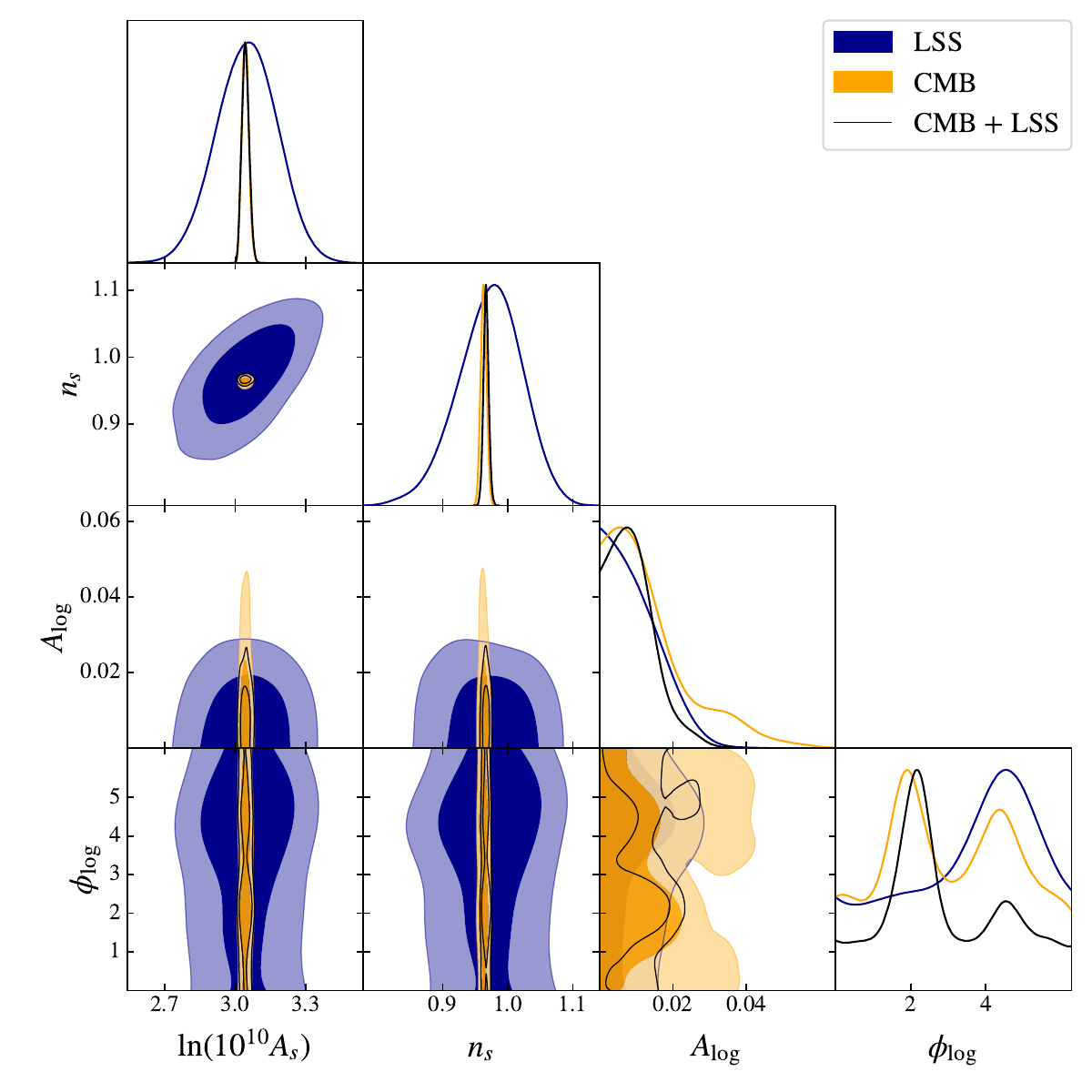}
    \caption{1D and 2D posterior distributions reconstructed from the LSS, CMB, and CMB + LSS datasets for the logarithmic oscillation analysis.}
    \label{fig:log_oscillations}
\end{figure}

\begin{figure}
    \centering
    \includegraphics[width=0.49\textwidth]{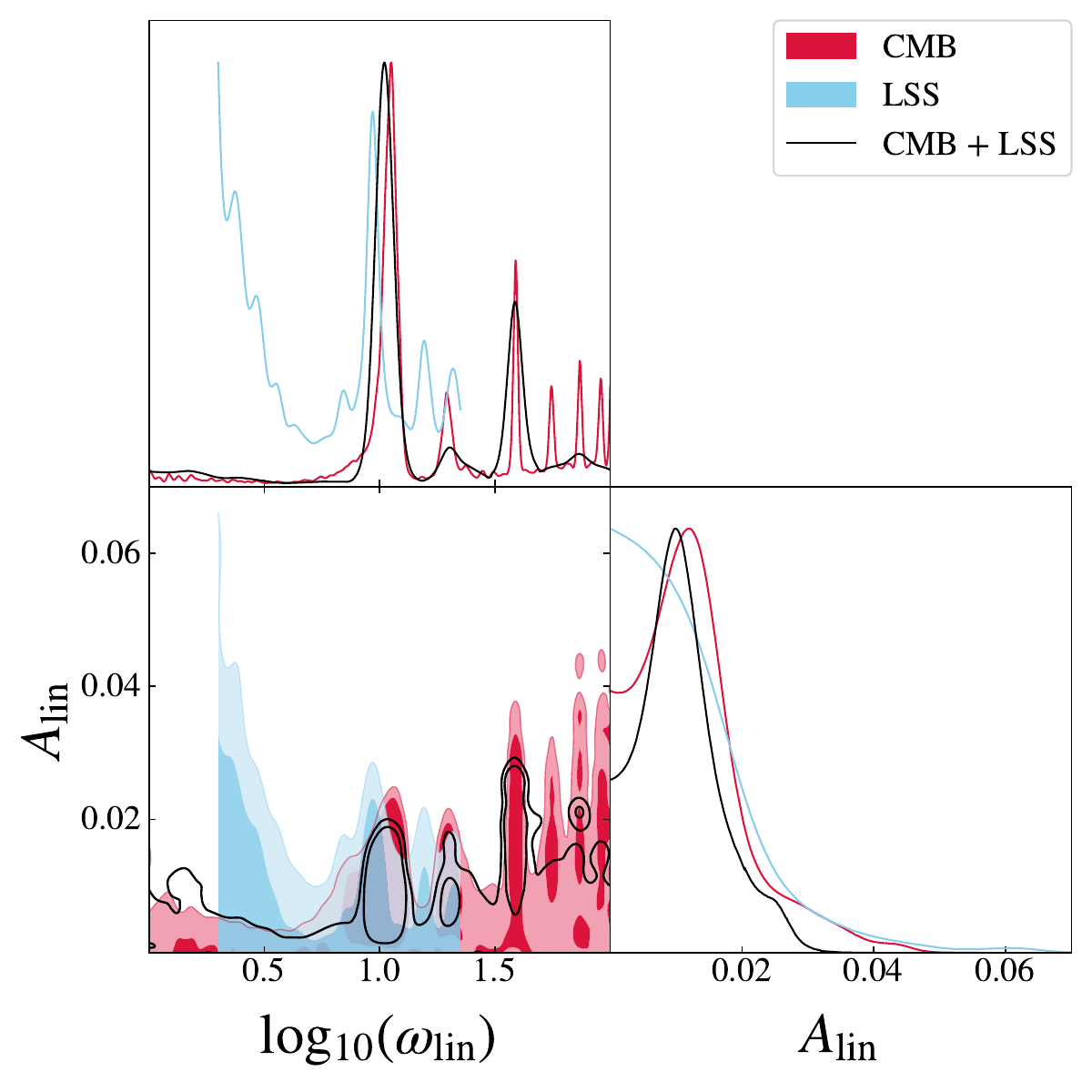}
    \includegraphics[width=0.49\textwidth]{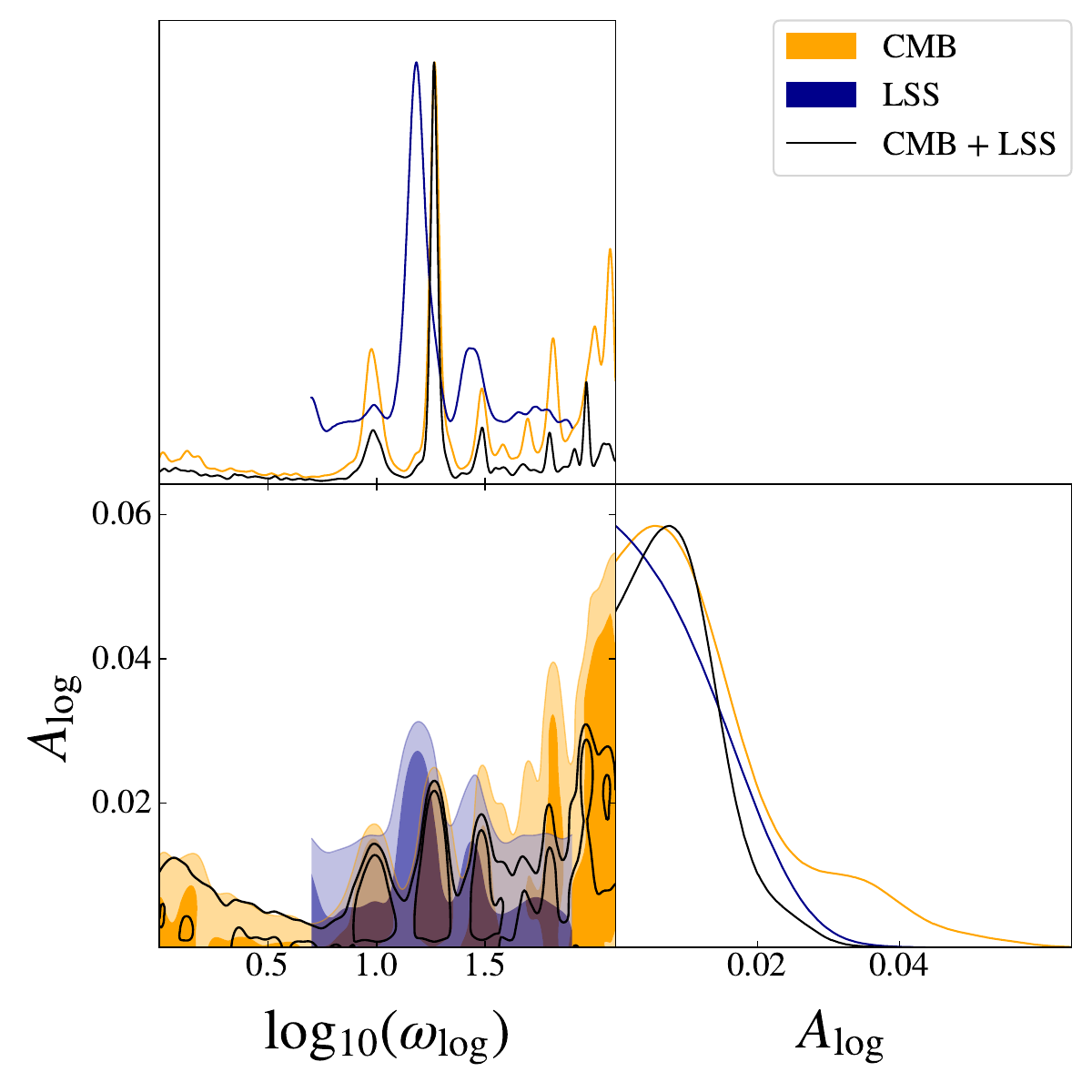}
    \caption{\textit{Left:} 1D and 2D posterior distributions of the $\log_{10}(\omega_{\rm lin})- A_{\rm lin}$ plan reconstructed from the CMB, LSS, and CMB + LSS analyses for the linear oscillations. \textit{Right:} Same for the logarithmic oscillations.}
    \label{fig:w_A}
\end{figure}

In \cref{fig:lin_oscillations,fig:log_oscillations}, we show our linear and logarithmic oscillation results from the LSS, CMB, and CMB + LSS datasets for the set of parameters $\{ \ln(10^{10} A_s), \, n_s, \, A_X, \, \phi_X \}$, while, in \cref{fig:w_A}, we display the $\log_{10}(\omega_{\rm X})- A_{\rm X}$ plan reconstructed from the same analyses. Let us note that we have checked that the CMB analysis applied to the two models studied here is consistent with Ref.~\cite{Planck:2018jri}.\\

As expected,  we can see in \cref{fig:lin_oscillations,fig:log_oscillations} that the CMB dataset largely dominates the constraints on $\ln 10^{10}A_s$ and $n_s$.
In addition, similarly to the CMB dataset, the LSS dataset is not capable of constraining $\omega_{\rm X}$ and $\phi_{\rm X}$ within the restricted priors on $\log_{10}(\omega_{\rm X})$.
However, the LSS data are able to provide strong constraints on the amplitude of linear and logarithmic oscillations $A_{\rm X}$. 
Note that when we compare the constraints from the CMB dataset with those obtained from the LSS dataset, we impose the same prior on $\log_{10}(\omega_{\rm X})$, namely the LSS prior (which is the most restrictive one), since the constraints on $A_{\rm X}$ depend on this prior, as can be seen in \cref{fig:w_A_LSS_priors} of \cref{app:lin_log}.
Therefore, imposing the LSS priors on $\log_{10}(\omega_{\rm lin})$, we find\footnote{The associated bestfit values are $A_{\rm lin}^{\rm LSS} = 0.013$ and $A_{\rm lin}^{\rm CMB}= 0.012$.}
\begin{align*}
    A_{\rm lin}^{\rm LSS} &< 0.031 \, , \\
    A_{\rm lin}^{\rm CMB} &< 0.019 \, ,
\end{align*}
at $95\%$ CL (see \cref{fig:w_A_LSS_priors} of \cref{app:lin_log} for a comparison between LSS and CMB with the LSS prior on $\log_{10}(\omega_{\rm X})$). The LSS constraint on the amplitude of the linear oscillations is of the same order of magnitude, and only $\sim 1.6$ times weaker than that of the CMB. For the logarithmic oscillation analysis, we obtain the same constraint on the logarithmic amplitude between the LSS and CMB analyses (when we impose the LSS prior on $\log_{10}(\omega_{\rm log})$), namely:\footnote{The associated bestfit values are $A_{\rm log}^{\rm LSS}= 0.023$ and $A_{\rm log}^{\rm CMB}= 0.018$.}
\begin{align*}
    A_{\rm log}^{\rm LSS}, A_{\rm log}^{\rm CMB} &< 0.024 \, ,
\end{align*}
at $95\%$ CL (see \cref{fig:w_A_LSS_priors} of \cref{app:lin_log} for a comparison between LSS and CMB with the LSS prior on $\log_{10}(\omega_{\rm X})$).
Let us note that we have checked that the constraints on $A_{\rm X}$ do not depend on whether the prior on $\omega_{\rm X}$ is linear or logarithmic, as shown in \cref{fig:lin_prior_lin_log} of \cref{app:lin_log}.
We conclude here that, although the constraints on the amplitude and the tilt of the primordial power spectrum are still largely dominated by the CMB data, the constraints on the amplitude of the oscillations are similar (or of the same order of magnitude) between the two analyses. Note that by probing larger frequencies (using a smaller $\Delta k$), the LSS constraints would potentially outperform CMB ones.\\

The LSS and CMB analyses are consistent, implying that we can combine them to improve the constraints on $A_{\rm X}$ (see \cref{fig:w_A}). Considering now the CMB prior on $\log_{10}(\omega_{\rm X})$, we obtain, for the linear oscillation analysis,\footnote{The associated bestfit values are $A_{\rm lin}^{\rm CMB} = 0.014$ and $A_{\rm lin}^{\rm CMB + LSS}= 0.013$.}
\begin{align*}
    A_{\rm lin}^{\rm CMB} &< 0.029 \, , \\
    A_{\rm lin}^{\rm CMB + LSS} &< 0.022 \, ,
\end{align*}
at $95 \%$ CL, corresponding to a $\sim 25 \%$ improvement of the CMB + LSS analysis over CMB alone. For the logarithmic oscillation analysis, we obtain,\footnote{The associated bestfit values are $A_{\rm log}^{\rm CMB} = 0.013$ and $A_{\rm log}^{\rm CMB + LSS} = 0.012$.}
\begin{align*}
    A_{\rm log}^{\rm CMB} &< 0.038 \, , \\
    A_{\rm log}^{\rm CMB + LSS} &< 0.021 \, ,
\end{align*}
at $95 \%$ CL, corresponding to a $\sim 50 \%$ improvement of the CMB + LSS analysis over CMB alone. Finally, the LSS dataset is very sensitive to (both linear and logarithmic) oscillatory features in the matter power spectrum, allowing us to significantly improve the constraints on the amplitude of the PPS oscillations compared with CMB alone.
In the following, we show that the LSS analysis can be even more constraining in the case of local features.

We finish this section by quantifying prior volume effects in our analysis. As discussed in \cref{Analysis}, an EFT analysis of the (e)BOSS data can be subject to prior volume effects.\footnote{We expect this to be even more pronounced in our analysis, as we recover the $\Lambda$CDM model when $A_X \to 0$. In this limit, the other PPS parameters can take any value, which increases the volume of the posterior distribution and favors (from a Bayesian point of view) small values of $A_X$.}
Using the metric defined in \cref{eq:projection_effect}, we find, for the linear oscillation scenario, that the distance of the mean (of the posteriors) from the bestfit is $n \sigma \lesssim 0.9 \sigma$ for LSS and  $n \sigma \lesssim 0.7 \sigma$ for CMB + LSS. For the logarithmic oscillation scenario, we find  $n \sigma \lesssim 1.3 \sigma$ for LSS and $n \sigma \lesssim 0.6 \sigma$ for CMB + LSS.\footnote{Note that we only consider the $\Lambda$CDM parameters to perform this exercise, since the additional PPS parameters are not detected.}  We note that this is within the expected range based on previous analyses in the $\Lambda$CDM context~\cite{Simon:2022lde,Holm:2023laa} (see \textit{e.g}, Refs.~\cite{Poudou:2025qcx,Lu:2025gki} in the context of beyond $\Lambda$CDM model), and that they noticeably decrease when we add \textit{Planck} on top of our LSS dataset. 
In addition, we note that the bestfit values of the amplitude $A_X$ are always roughly located in the $68 \%$ CL, except for the LSS analysis of the logarithmic oscillations (where projection effects are most important).
We explore these projection effects in more detail below.

\section{Local features: two case studies}\label{sec:one_spec}

As discussed previously, many well-motivated ultraviolet (UV) completions of inflation naturally give rise to richer scenarios, involving multiple fields and non-canonical kinetic terms. These more complex models often predict sharp or localized primordial features that break the near scale-invariance of the power spectrum \cite{Starobinsky:1992ts,Adams:2001vc,Achucarro:2010da,Gao:2012uq,Gao:2013ota,Noumi:2013cfa,Achucarro:2013cva,Braglia:2020fms,Braglia:2021ckn,Braglia:2021sun,Braglia:2021rej,Braglia:2022ftm}. For example, sharp variations in the inflaton sound speed typically lead to damped linear oscillatory features \cite{Chen:2011zf,Achucarro:2010da}.
Introducing such features in the PPS can also modify the inferred background cosmological parameters, including the physical baryon density $\omega_b = \Omega_b h^2$, and the Hubble constant $H_0$ (see \textit{e.g.}, \cite{Kinney:2001js,Keeley:2020rmo}). Given their ability to impact key cosmological parameters, primordial features have recently been explored as a possible explanation for some of the observational challenges plaguing \lcdm~ \cite{Bull:2015stt,Perivolaropoulos:2021jda,Abdalla:2022yfr}. A particularly intriguing example is the lensing anomaly observed in the \planck\ (PR3) CMB data \cite{Planck:2018lbu,Motloch:2018pjy} (see also the recent lensing measurement of \cite{SPT-3G:2024atg}), which exhibits a known degeneracy with local primordial features \cite{Hazra:2014jwa,GallegoCadavid:2016wcz,Domenech:2019cyh,Domenech:2020qay,Ballardini:2022vzh}. While the origin of this anomaly\textemdash statistical, physical or systematic\textemdash remains under discussion, the possibility of a primordial origin is especially compelling, as it would point to non-trivial inflationary dynamics beyond the standard slow-roll paradigm.

In this section, we applied our LSS analysis to two PPS exhibiting localized features: a first one with localized linear oscillations, thanks to the ``\textit{One Spectrum}'' template, and a second one with localized logarithmic oscillations, thanks to the ``damped logarithmic oscillations'' model.

\subsection{One spectrum-like template}
\begin{table}[t]
    \centering
    \begin{tabular}{lcc}
    \hline\hline
     Model & Parameter & Prior\\  
    \hline
    \multirow{4}{*}{One Spectrum} & $\alpha$ &  $\mathcal{U}[0, 0.2]$  \\
    & $\log_{10} (\beta \ /[\rm Mpc^{4}])$ &  $\mathcal{U}[2, 8]$  \\
    & $\omega \ [\rm Mpc]$ &  $\mathcal{U}[240, 400]$  \\
    & $k_0 \ [\rm Mpc^{-1}]$ &  $\mathcal{U}[0.13, 0.14]$  \\
    \hline\hline
    \end{tabular}
    \caption{Priors used in the \textit{One Spectrum} analysis \cite{Hazra:2022rdl}. }
    \label{tab:OneSpectrum_priors}
\end{table}

We first focus on the so-called \textit{One Spectrum} \cite{Hazra:2022rdl}, a PPS that was reconstructed to address the lensing ($A_\mathrm{L}$)\footnote{The degeneracy between lensing anomaly and local primordial features was first shown in~\cite{Hazra:2014jwa}.} and curvature ($\Omega_k$) anomalies in the \planck\ (PR3) data, while alleviating the $S_8$ and $H_0$ tensions~\cite{Motloch:2018pjy,DiValentino:2019qzk} to certain extent. For simplicity, we work with the analytical template used in \cite{Hazra:2022rdl}\footnote{Similar analytical form was introduced in~\cite{Planck:2018jri} to mimic the lensing anomaly.} that closely mimics the {\it One Spectrum} PPS:
\begin{equation}\label{eq:onespectrum}
    \mathcal{P}_\zeta(k)=\mathcal{P}_{\zeta,0}(k)\left[1+\frac{\alpha\sin(\omega(k-k_0))}{1+\beta(k-k_0)^4}\right]~,
\end{equation}
where $\mathcal{P}_{\zeta,0}$ is the standard power-law given by \cref{eq:PL}, and where $\alpha, \ \beta, \ \omega$ and $k_0$ are free parameters of the model. Note that the standard power-law form is recovered in the limit $\alpha\to0$. The priors used for the parameters of this model match those of Ref.~\cite{Hazra:2022rdl}, and are reported in \cref{tab:OneSpectrum_priors} for completeness. In \cref{sec:priors}, we show that the priors used here, namely $\omega \in [240, 400]$ Mpc, are within the frequency sensitivity regime of our analysis.
Note that in the linear oscillation analysis (see \cref{sec:priors}), unlike in the \textit{One Spectrum} analysis, we normalize the wavenumber $k$ by $k_*$, implying that the \textit{One Spectrum} sensitivity regime is $\omega \in [2, 22.5] / k_* = [40, 450]$ Mpc, which largely includes the prior used in \cref{tab:OneSpectrum_priors}.

\begin{figure}
    \centering
    \includegraphics[width=1\linewidth]{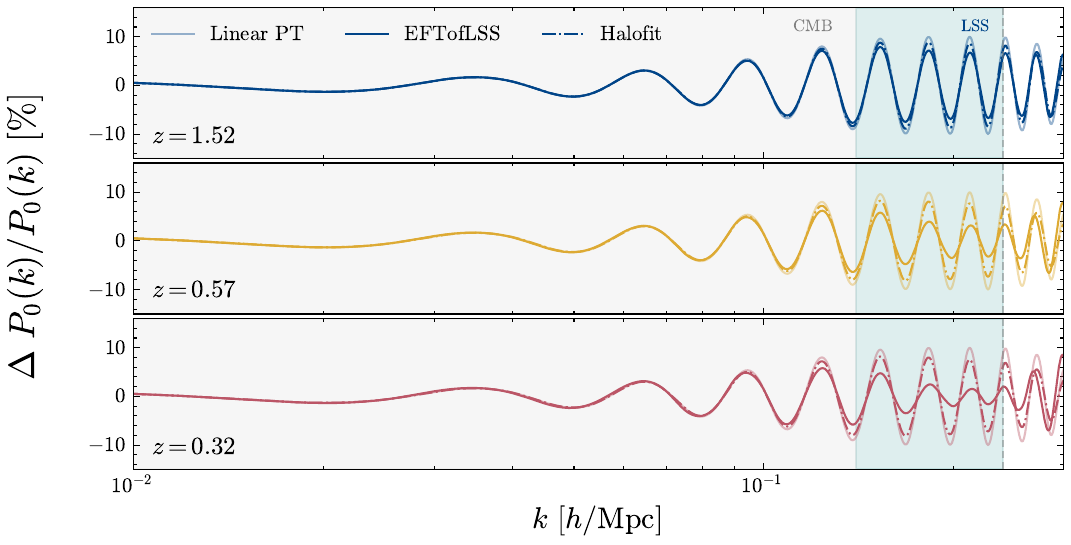}
    \caption{Non-linear matter power spectrum residuals of the \OSpec template (see \cref{eq:onespectrum}) with respect to a featureless power-law, considering an amplitude $\alpha=0.1$ and a frequency  $\omega=315$ Mpc. We show the prediction from the linear perturbation theory, the EFTofLSS as well as from \texttt{Halofit}, respectively.
    From top to bottom, we consider the eBOSS, BOSS CMASS and BOSS LOWZ effective redshifts, respectively, illustrating how non-linear structure formation processes wash away oscillatory features in the power spectrum.} 
    \label{fig:OS_mPk_residuals}
\end{figure}

In \cref{fig:OS_mPk_residuals}, we plot the residuals of the non-linear matter power spectrum (with respect 
to the standard power law) for the effective redshift bins included in our analysis.
As in the case of global linear and logarithmic oscillations (see \cref{fig:Nbody-comp,fig:Nbody-comp-lin}), we observe that oscillatory features in the power spectrum are progressively washed out by gravitational non-linearities, with the damping effect becoming more pronounced at later times. 

\subsubsection{Cosmological constraints}\label{sec:res_onespec}

\begin{figure}
    \centering
    \includegraphics[width=0.7\textwidth]{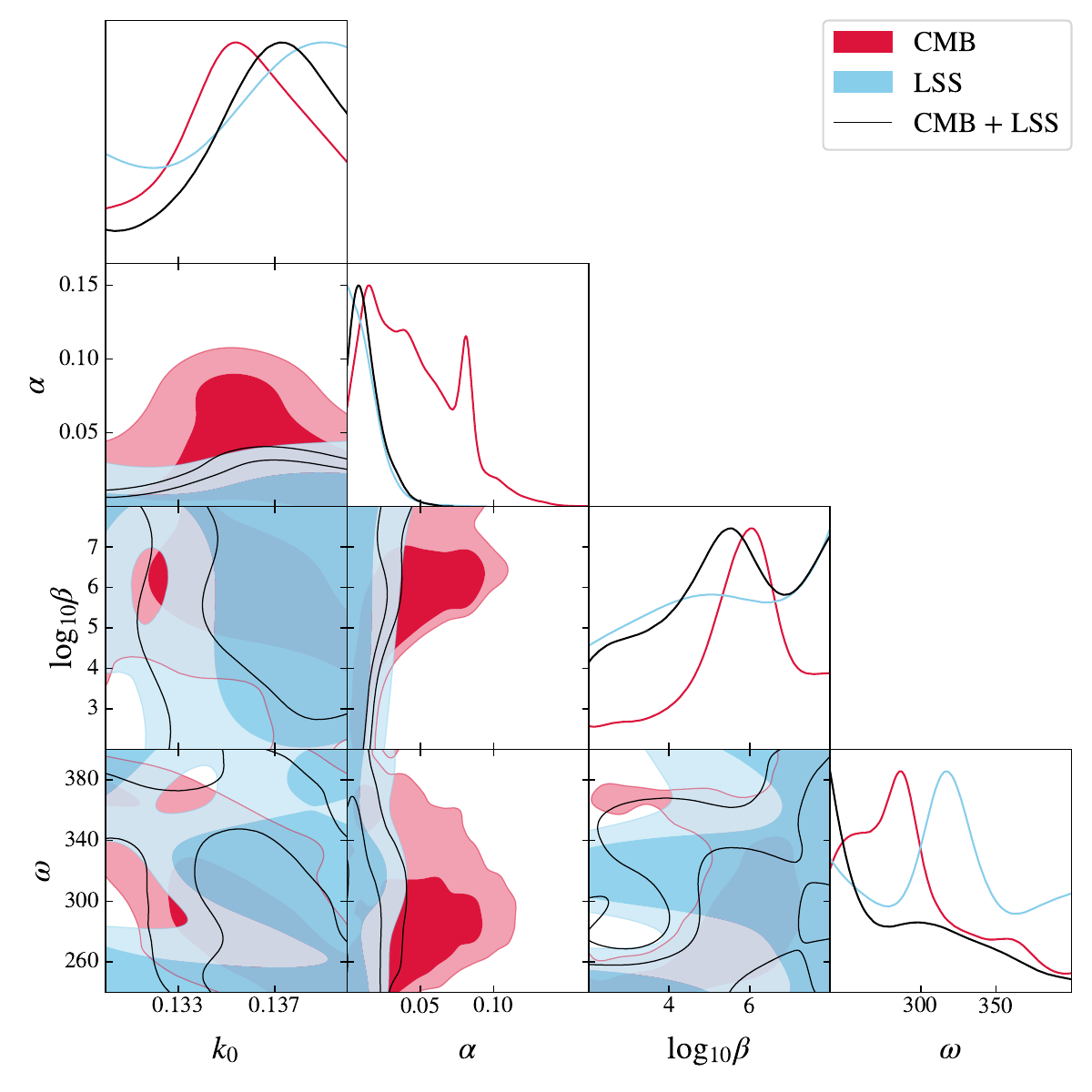}
    \caption{1D and 2D posterior distributions reconstructed from the CMB, LSS and CMB + LSS datasets for the \OSpec analysis.}
    \label{fig:OneSpectrum}
\end{figure}

In \cref{fig:OneSpectrum}, we show our results for the \OSpec analysis from the LSS, CMB and CMB + LSS datasets.
As previously, we can see that CMB data largely dominate the constraints on $\ln 10^{10}A_s$ and $n_s$ compared with LSS data. In addition, the parameters $k_0$, $\log_{10} \beta$ and $\omega$ are not detected in LSS data, as in CMB data.
However, the LSS data are $\sim 3$ times more constraining on $\alpha$, the amplitude of the \textit{One Spectrum} oscillations, than the CMB data, with
\begin{align*}
    \alpha^{\rm CMB} &< 0.094 \, , \\
    \alpha^{\rm LSS} &< 0.034 \, , \\
    \alpha^{\rm CMB + LSS} &< 0.035 \, ,
\end{align*}
at $95 \%$ CL.
When combining the two datasets, we obtain a tight constraint on this parameter, mainly coming from LSS data.

\begin{figure}
    \centering
    \includegraphics[trim=0.8cm 0cm 3cm 1.5cm, clip=true,width=1.\columnwidth]{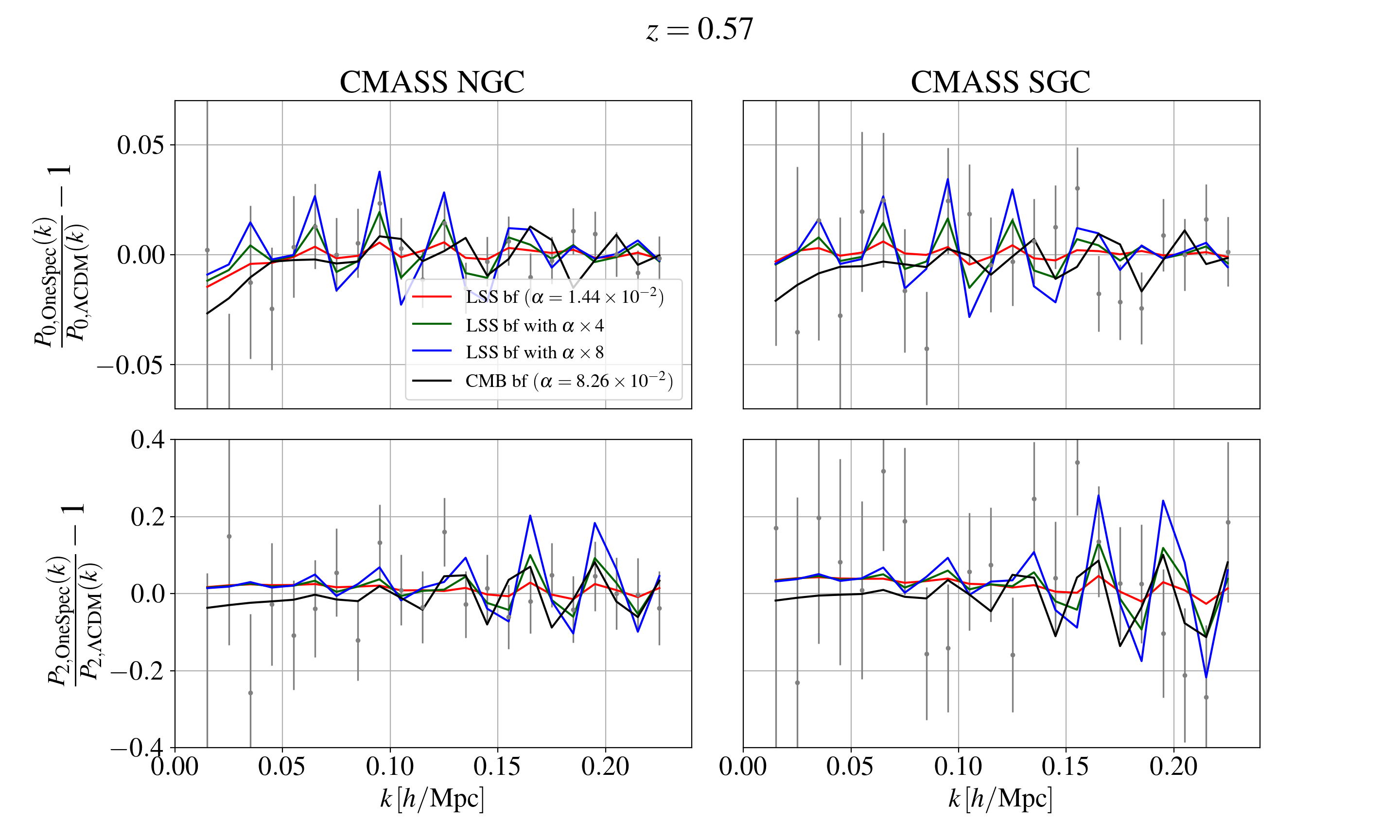}

    \begin{tabular}{c|cccc|c}
    \hline\hline
        Data & BOSS CMASS & BOSS LOWZ & eBOSS & BBN & Total \\
    \hline
       $\chi^2_{\rm LSS}$ & $81.02$ &   $74.43$ & $59.36$ & $0.71$ & $215.52$ \\
    \hline
       $\chi^2_{\rm CMB}$ & $109.10$ &   $88.49$ & $59.53$ & $ 0.24$ & $257.37$ \\
    \hline\hline
    \end{tabular}
    \caption{ Residuals of the monopole and quadrupole of the galaxy power spectra with respect to the $\Lambda$CDM model for the two sky cuts of the BOSS CMASS sample. The red line corresponds to the LSS bestfit, while the green and blue lines correspond to the LSS bestfit with $\alpha$ multiplied by 4 and 8, respectively. Finally, the black line corresponds to the \textit{Planck} bestfit, where we minimize the EFT parameters. In \cref{fig:residuals_OneSpectrum_eBOSS}, we display the same figure for the BOSS LOWZ and eBOSS samples.
    In the table, we display the $\chi^2$ of the LSS and CMB bestfits when applied to the LSS dataset.}
    \label{fig:residuals_OneSpectrum_BOSS}
\end{figure}

To better understand where the LSS constraining power on this model is coming from, we plot in \cref{fig:residuals_OneSpectrum_BOSS} the residuals of the monopole and quadrupole of the BOSS CMASS galaxy power spectra with respect to $\Lambda$CDM. The same figure is shown for the BOSS LOWZ and eBOSS QSO samples in \cref{fig:residuals_OneSpectrum_eBOSS} of \cref{app:power_spectra_one_spectrum}.
In these figures, we represent the LSS bestfit, the LSS bestfit with $\alpha$ multiplied by 4 and 8, as well as the CMB bestfit. 
For the latter, we set the cosmological parameters to the CMB bestfit, and we minimize the EFT parameters accordingly.
In addition, in the table of \cref{fig:residuals_OneSpectrum_BOSS}, we display the $\chi^2$ of the LSS and CMB bestfits when applied to the LSS dataset.
Interestingly, we can see that the CMB bestfit is excluded by our LSS dataset at more than $6 \sigma$, with $\Delta \chi^2 \equiv \chi^2_{\rm CMB} - \chi^2_{\rm LSS} = + 41.9 $.
This suggests that there is no degeneracy between the EFT parameters and the cosmological parameters that could loosen the LSS constraints.
The exclusion of the CMB bestfit mainly comes from the scales above $k \sim 0.16 h {\rm Mpc}^{-1}$ of the  BOSS CMASS and BOSS LOWZ samples, where the CMB oscillations are out of phase of the LSS ones and where the errors of the data are smaller. In addition, when we increase the $\alpha$ parameter, the amplitude of the oscillations quickly becomes too large for the clustering data (see \cref{fig:residuals_OneSpectrum_BOSS}), explaining the strong constraint obtained on the amplitude of oscillations in our analysis. We have verified that when considering only the monopole\textemdash rather than both the monopole and quadrupole as in our baseline analysis\textemdash the constraints on $\alpha$ are relaxed by a factor of 2. This suggests that a significant portion of the constraining power arises from the degraded fit to the quadrupole, especially at high-$k$, as illustrated in \cref{fig:residuals_OneSpectrum_BOSS,fig:residuals_OneSpectrum_eBOSS}.
Finally, this analysis shows that an EFTofLSS analysis applied to (e)BOSS data is able to disfavor localized primordial oscillations that are in fact favored by \textit{Planck}.
We expect this conclusion to be applicable to other models with localized primordial oscillations located within the regime of validity of the EFTofLSS.

\begin{figure}
    \centering
    \includegraphics[width=\textwidth]{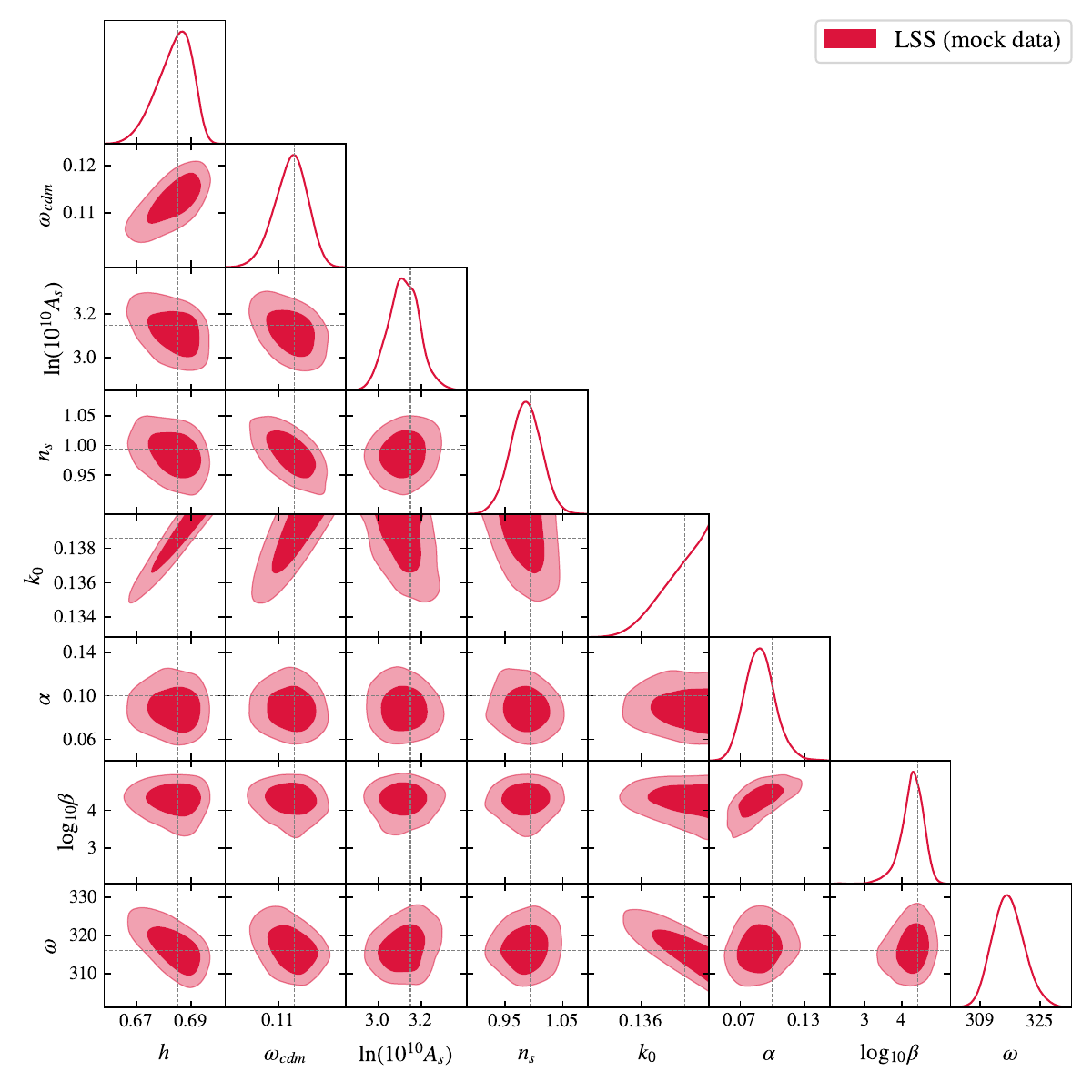}
    
    \begin{tabular}{ccccccccc}
    \hline\hline
        $\alpha$ &$\log_{10}\beta$ & $\omega$ & $k_0$ & $\ln 10^{10} A_s$ & $n_s$ &$h$& $10^2\omega_b$  & $\omega_{c}$  \\
    \midrule
      $0.1$ & $4.44$ &   $315.9$ & $0.1386$ & $3.150$ & $0.9937$ & $0.6853$ & $2.243$ & $0.1133$ \\
      \hline\hline
    \end{tabular}
    
    \caption{1D and 2D posterior distributions reconstructed from BOSS and eBOSS mock data generated with the LSS bestfit, together with $\alpha = 0.1$. The dashed lines represent the truth of the mock data, which is displayed in the table.}
    \label{fig:OneSpectrum_mock}
\end{figure}

\subsubsection{Validation with Mocks}\label{sec:mocks}

In this subsection, we assess the robustness of our results thanks to mock data.
In \cref{fig:OneSpectrum_mock}, we fit BOSS and eBOSS mock data using the same priors and the same (e)BOSS covariance matrices as for the real data analysis.
The truth of the mock data corresponds to the LSS bestfit (reported in the table of \cref{fig:OneSpectrum_mock}), together with $\alpha = 0.1$, a value which is excluded at $\sim 6 \sigma$ by the LSS dataset and which lies on the $2 \sigma$ limit of the CMB dataset.
In \cref{fig:OneSpectrum_mock}, we can see the strong constraining power of the LSS data on this particular PPS parametrization, since we obtain, with LSS only, a strong detection of this model.
Indeed, we reconstruct $\alpha = 0.089^{+0.013}_{-0.015}$  at $68\%$ CL, corresponding to a $15\%$ precision detection of this parameter. This analysis therefore corroborates the strong constraining power of the real data analysis as well as the features we see in the (e)BOSS galaxy power spectra.\\

In addition, this mock exercise allows us to gauge the impact of prior volume projection effect on the cosmological results for one of the most complex models of our analysis.
Following the metric defined in \cref{eq:projection_effect}, we find that the distance of the mean (of the posteriors) from the truth is $n \sigma \lesssim 0.4 \sigma$ for all cosmological parameters, except for $\alpha$ and $\log_{10} \beta$ where $ n \sigma  \sim 0.8 \sigma$ and $n \sigma \sim 0.6 \sigma$, respectively, showing that the projection effects are rather small in our analysis (given the large number of free parameters). 
In order to cross-check this statement, we display in \cref{fig:One_spectrum_bestfit} of \cref{app:power_spectra_one_spectrum} the 1D and 2D posterior distributions of the LSS analysis (with real data), together with the associated bestfit values. 
The distance of the mean from the bestfit is equal to $0.3 \sigma$, $0.5 \sigma$, $0.1 \sigma$, $1.2 \sigma$ and $0.7 \sigma$ on $h$, $\omega_{\rm cdm}$, $\omega_b$, $\ln 10^{10} A_s$, and $n_s$, respectively (the other parameters not being detected).
Given the large number of EFT and cosmological parameters in our analysis and according to Ref.~\cite{Simon:2022lde}, these shifts are rather acceptable ($\lesssim1\sigma$ for all parameters, except for $\ln{(10^{10} A_s)}$) and confirm the robustness of our analysis.

\subsection{Damped logarithmic oscillations}

\begin{figure}
    \centering
    \includegraphics[width=\linewidth]{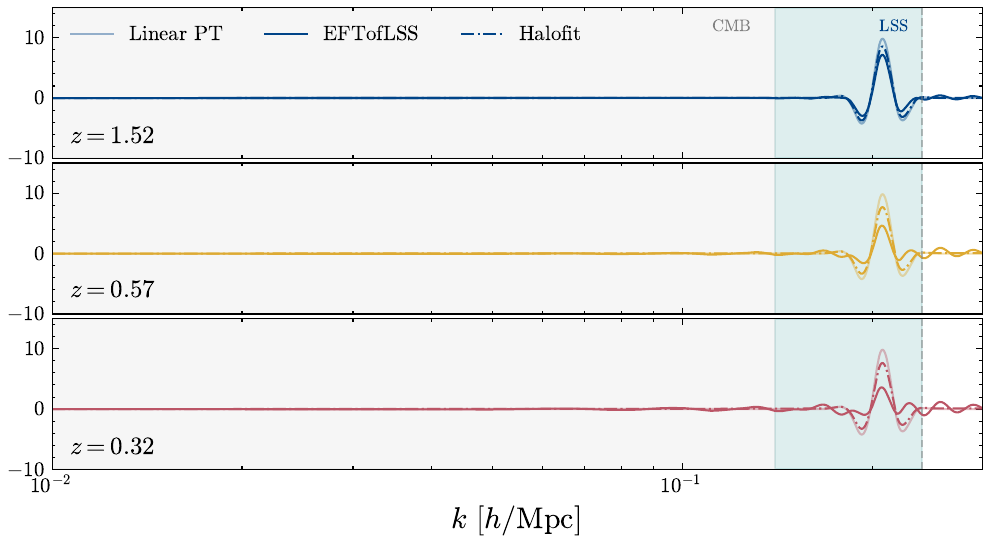}
    \caption{Non-linear matter power spectrum residuals for the damped logarithmic oscillation template (see \cref{eq:log-osc-damped}) with respect to a featureless power-law, considering an amplitude $A_{\rm log}=0.1$, a frequency  $\log_{10}\omega_{\rm log}=1.55$, $\mu=0.14$ and $\beta=6$. We show the prediction from the linear perturbation theory, the EFTofLSS as well as from \texttt{Halofit}, respectively.
    From top to bottom, we consider the eBOSS, BOSS CMASS and BOSS LOWZ effective redshifts, respectively.}
    \label{fig:Dlog_mPk_residuals}
\end{figure}

We turn our attention to a second  example of local features, proposed in Ref.~\cite{Antony:2024vrx}, which is a generalization of the global logarithmic oscillation model:
\begin{equation}\label{eq:log-osc-damped}
\mathcal{P}_\zeta(k)=\mathcal{P}_{\zeta,0}(k)\left[1+A_{\rm log}\cos\left(\omega_{\rm log}~\frac{k}{k_*} +\phi_{\rm log}\right) e^{-\frac{\beta^2(k-\mu)^2}{2k_*^2}}   \right]~,
\end{equation}
where $\mu$ and $\beta$ are respectively the center and the width of the Gaussian envelope. In this model, the phase is determined by $\mu$ and $\beta$ in such a way that the peak of the Gaussian is located at a maximum of the oscillations: $\phi_{\rm log} = - \omega_{\rm log}  \ln(\mu/k_*)$. 
Let us note that when $\beta \to 0$, we recover the global logarithmic oscillations explored in \cref{sec:Global}.
Following Ref.~\cite{Antony:2024vrx}, we impose the following priors: $A_{\rm log} \in [0, 0.5]$, $\log_{10}(\omega_{\rm log}) \in [1.0, 1.8]$, $\mu \in [0.001, 0.175] \ {\rm Mpc}^{-1}$  and $\beta \in [0,10]$. 
According to the methodology developed in \cref{sec:priors}, we can safely use these priors in the LSS analysis.
In \cref{fig:Dlog_mPk_residuals}, we plot the non-linear matter power spectrum residuals for the effective redshifts considered in our analysis. We note that for large $\beta$, \textit{i.e.}, very localized oscillatory features, the constraints on the amplitude $A_{\rm log}$ can be significantly relaxed with respect to the global logarithmic oscillations, as for the choice of cosmological parameters shown in \cref{fig:Dlog_mPk_residuals}.

\begin{figure}
    \centering
    \includegraphics[width=0.7\textwidth]{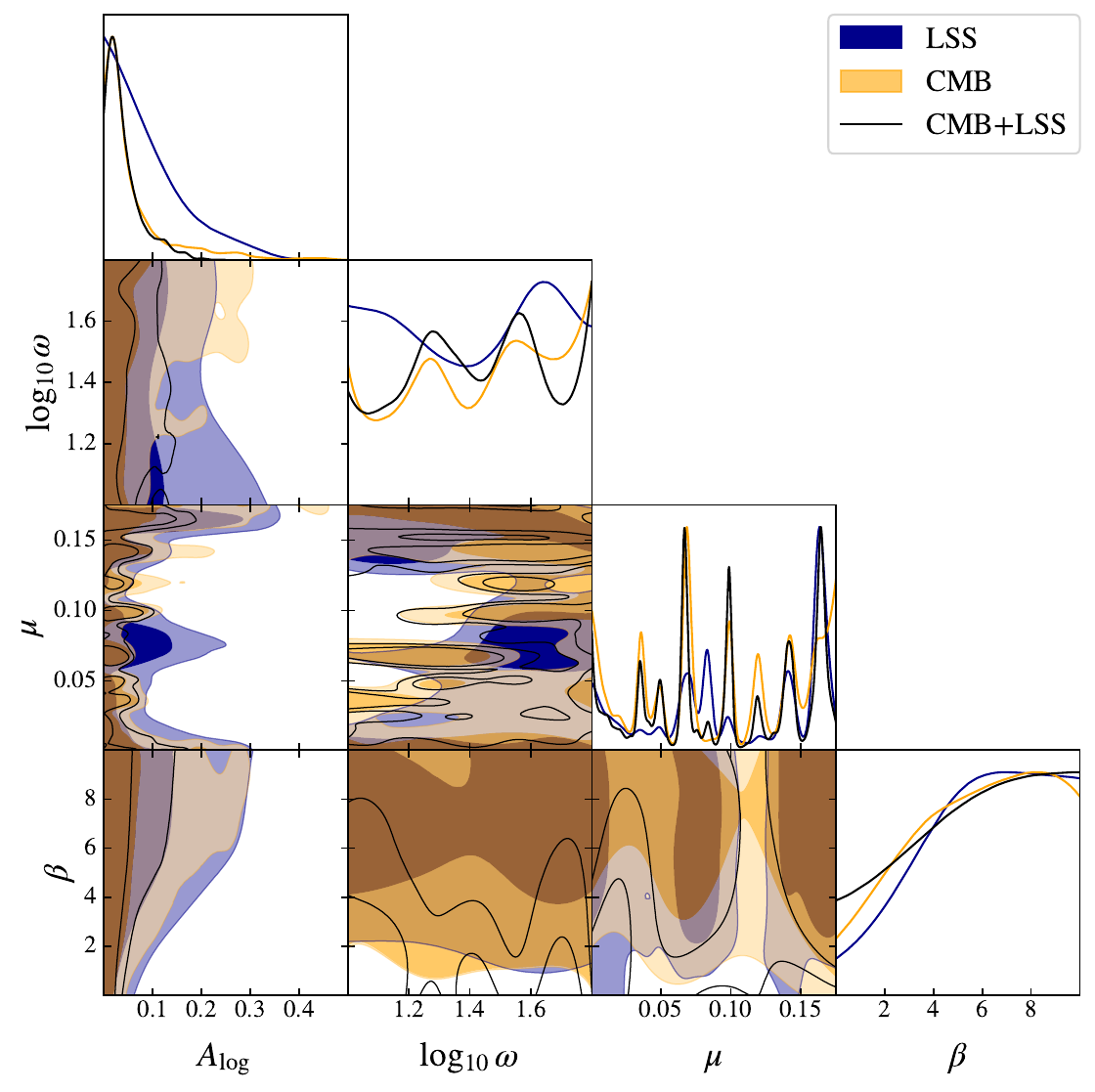}
    \caption{1D and 2D posterior distributions reconstructed from the LSS and CMB datasets for the damped logarithmic oscillation analysis.}
\label{fig:dampedlog_oscillations}
\end{figure}

In \cref{fig:dampedlog_oscillations}, we display the 1D and 2D posterior distributions reconstructed from the LSS and CMB datasets for the damped logarithmic oscillation analysis.
In particular, for the amplitude of the oscillations, we obtain at $95 \%$ CL
\begin{align*}
    A_{\rm log}^{\rm LSS} &< 0.262 \, , \\
    A_{\rm log}^{\rm CMB} &< 0.237 \, , \\
    A_{\rm log}^{\rm CMB+LSS} &< 0.124 \, . \\
\end{align*}
Once again, we can see that LSS constraints alone are competitive and complementary to the CMB constraints, and offer a promising way of tightly constraining the amplitude of primordial features with the next generation of LSS surveys. 

\section{Is the reconstructed PPS from CMB allowed by LSS?}\label{sec:regMRL}

\subsection{Deconvolved PPS from CMB data}

Deriving constraints on the very early universe usually amounts to writing down a potential for the inflaton and computing the corresponding power spectrum of curvature fluctuations.
An alternative approach, driven by the data, involves deconvolving the observed CMB power spectra $C_\ell$'s to infer the corresponding PPS  given an assumed background cosmology (\textit{i.e.}, a specific transfer function) \cite{Hannestad:2000pm,Shafieloo:2003gf,Hazra:2014jwa,Planck:2015sxf,Chandra:2021ydm,Hazra:2022rdl,Sohn:2022jsm,Lodha:2023jru}. The observed $C_\ell$'s are related to the PPS $P_\zeta(k)$ via
\begin{equation}\label{eq:cls}
C_\ell^{XY} = 4\pi \int d \ln{k} ~\Delta^X_\ell(k)~\Delta^Y_\ell(k)~ P_\zeta(k)~,
\end{equation}
where $\Delta_\ell^{X}(k;\vect{\theta}_{\rm cosmo})$ represents the transfer function, computed using a Boltzmann solver for a given set of cosmological parameters $\vect{\theta}_{\rm cosmo}$, with $X = \{T, E\}$ denoting temperature and polarization, respectively. Considerable efforts have been dedicated to reconstructing the PPS directly from data \cite{Hannestad:2000pm,Bridle:2003sa,Sohn:2022jsm,Shafieloo:2003gf,Covi:2006ci,Shafieloo:2006hs,Bridges:2008ta,2009JCAP...07..011N,Hazra:2013xva,Chandra:2021ydm,Lodha:2023jru,Chaki:2025qsc,Raffaelli:2025kew}. However, as a proof of concept, we focus on the PPS reconstruction from \cite{Sohn:2022jsm}, which employs the \textit{regularized} Modified Richardson-Lucy (regMRL) algorithm \cite{Richardson:72,1974AJ.....79..745L,2014OptLT..58..100C}. The resulting form of the PPS was found to simultaneously improve the fit to the TT, TE, and EE \planck\ PR3 data \cite{Sohn:2022jsm}. 
We can assess the impact of this reconstruction on late-time observables by supplying the reconstructed PPS in tabulated format ($k, P_\zeta(k)$) to the Boltzmann solver \classy\ via the ``\texttt{external\_Pk}'' module. 
Thus, the reconstructed PPS, together with the background parameters $\vect{\theta}_\mathrm{cosmo}$, fully specifies the linear matter power spectrum (and the growth factor $f$) at late times. We use this as input to compute the non-linear galaxy power spectrum in redshift space, following the methodology outlined in \cref{Analysis}. By fitting this model to LSS data, we can compare inferred constraints on $\vect{\theta}_\mathrm{cosmo}$, serving as a crucial consistency test for the assumed background cosmology as well as for the initial conditions. Any discrepancy between the inferred values of $\vect{\theta}_\mathrm{cosmo}$ from LSS and those used for the CMB-based PPS reconstruction would indicate issues either with: (i) the assumed background cosmology, (ii) the reconstructed PPS, or (iii) unaccounted-for systematics in the data.

\subsection{LSS constraints}

\begin{figure}
    \centering
    \includegraphics[width=0.5\textwidth]{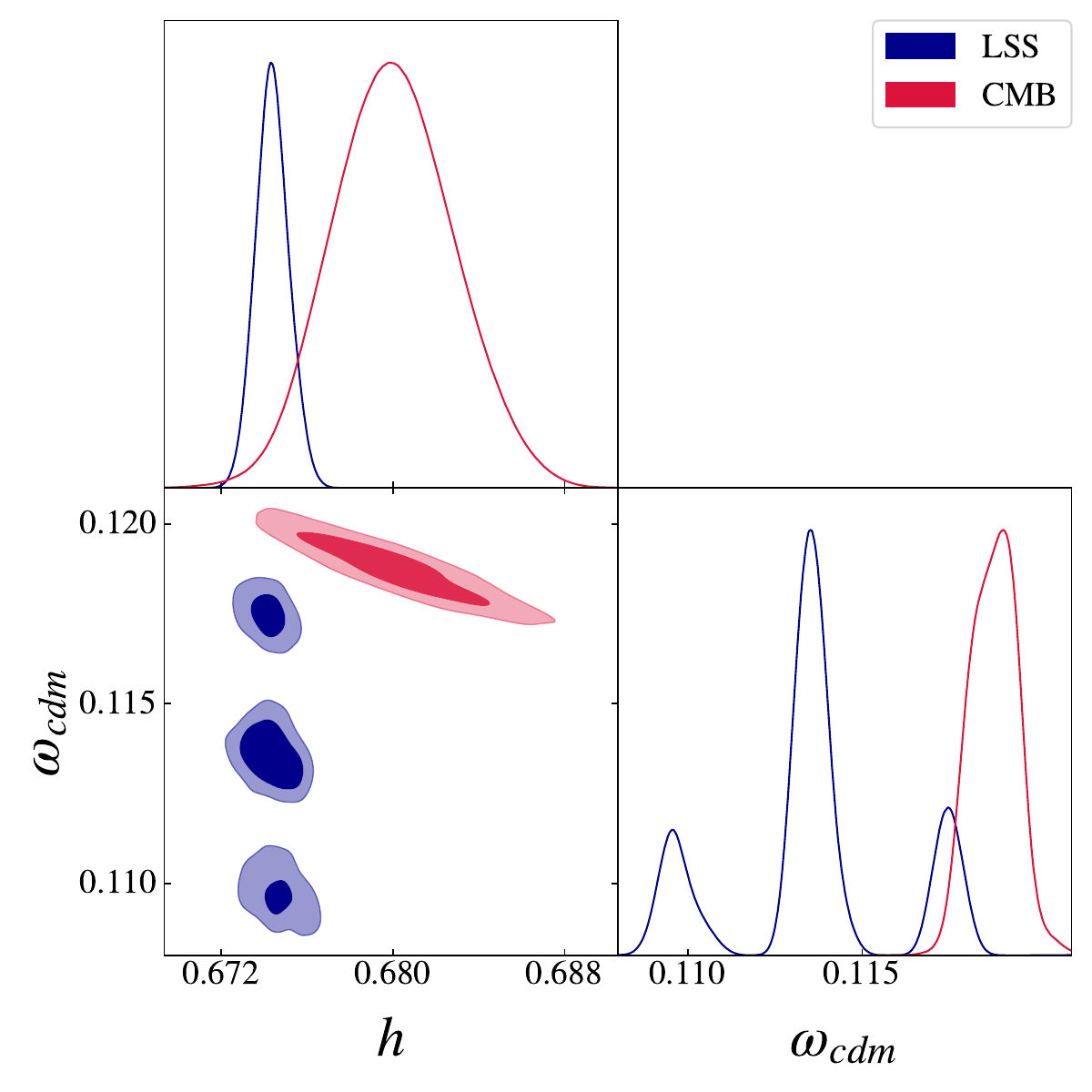}
    \caption{1D and 2D posterior distributions reconstructed from the LSS and CMB datasets for the regMRL analysis.}
    \label{fig:MRL}
\end{figure}

\begin{figure}
    \centering
    \includegraphics[trim=0.8cm 0cm 3cm 1.5cm, clip=true,width=1.\columnwidth]{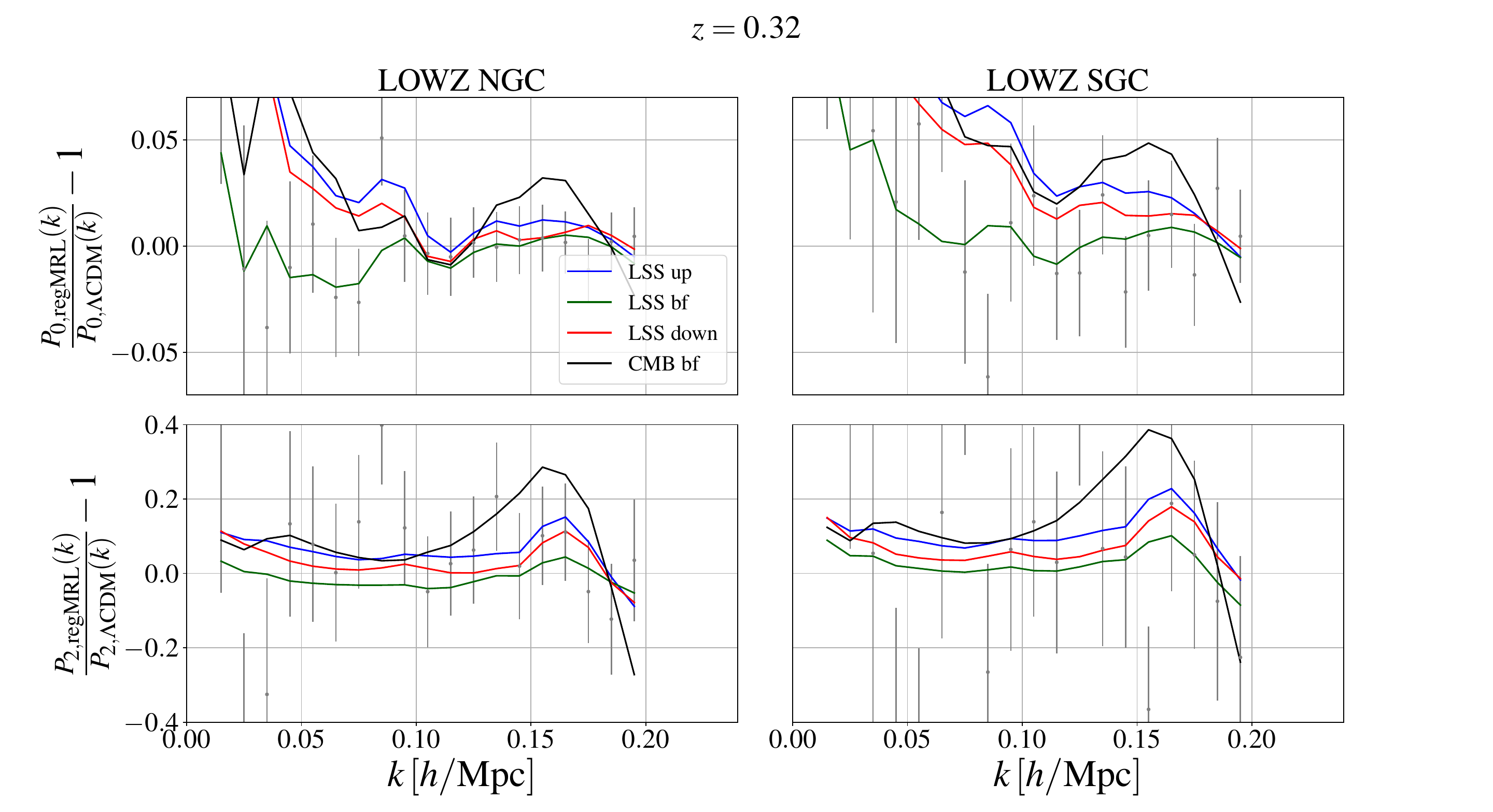}
    \caption{Residuals of the monopole and quadrupole of the galaxy power spectra with respect to the $\Lambda$CDM model for the two sky cuts of the BOSS LOWZ sample. The green line corresponds to the LSS bestfit (which lies in the middle posterior of \cref{fig:MRL}), while the blue and red lines correspond to the LSS bestfit with $\omega_{\rm cdm} = 0.1095$ (which lies in the bottom posterior of \cref{fig:MRL}) and the LSS bestfit with $\omega_{\rm cdm} = 0.1175$ (which lies in the upper posterior of \cref{fig:MRL}), respectively. Finally, the black line corresponds to the \textit{Planck} bestfit.}
    \label{fig:residuals_MRL}
\end{figure}

\cref{fig:MRL} presents our results for the regMRL analysis, comparing constraints from LSS and CMB datasets. Notably, the LSS posteriors exhibit a discrepancy with the CMB posteriors at more than $2\sigma$. This demonstrates that an EFTofLSS analysis of (e)BOSS data can challenge a given PPS reconstruction from \planck\ data.
Furthermore, \cref{fig:MRL} reveals a multi-modal posterior, suggesting that from an LSS perspective, certain primordial features can be mimicked by variations in $\omega_\mathrm{cdm} = \Omega_\mathrm{cdm} h^2$. To explore this, \cref{fig:residuals_MRL} shows the residuals of the BOSS LOWZ\footnote{Note that we observe the same trend for BOSS CMASS and eBOSS QSO samples, but we display BOSS LOWZ for the sake of clarity.} galaxy power spectra (relative to $\Lambda$CDM) for the LSS bestfit model (which lies in the middle mode of the posterior in \cref{fig:MRL}), the LSS bestfit with $\omega_{\rm cdm} = 0.1095$ (which lies in the lower mode of the posterior in \cref{fig:MRL}), and the LSS bestfit with $\omega_{\rm cdm} = 0.1175$ (which lies in the upper mode of the posterior in \cref{fig:MRL}). For the last two cases, we fix the cosmology but minimize the EFT parameters. The resulting $\chi^2$ values are: $\chi^2_{\rm LSS} = 227.9$ for the LSS bestfit, $\chi^2_{\rm LSS} = 229.3$ for $\omega_{\rm cdm} = 0.1175$, and $\chi^2_{\rm LSS} = 239.2$ for $\omega_{\rm cdm} = 0.1095$. 
The $\chi^2_{\rm LSS}$ is rather similar for these three values of $\omega_{\rm cdm}$, which explains the multimodal posterior.
Additionally, the CMB bestfit, for which we minimize the EFT parameters accordingly, is significantly disfavored by LSS data, with $\chi^2_{\rm LSS} = 290.1$. This suggests that if the early universe imprinted such features on the PPS, the galaxy power spectrum measurements would necessitate modifications to background parameters that would be in tension with CMB constraints. However, we emphasize that the regMRL PPS was reconstructed assuming the \planck\ bestfit, $\vect{\theta}_\mathrm{cosmo}=\vect{\theta}_{\rm Planck}^{\rm bf}$ \cite{Sohn:2022jsm}. 
Allowing $\vect{\theta}_\mathrm{cosmo} \neq \vect{\theta}_\mathrm{Planck}^{\rm bf}$ at the level of the CMB deconvolution could yield a slightly different reconstructed PPS, potentially reconciling CMB and LSS constraints. This highlights the potential for a joint reconstruction of PPS using both datasets, an avenue we leave for future work.

Assuming the LSS dataset is free from significant systematics, these results suggest that the improved fit from the regMRL features may stem from the algorithm fitting noise or unaccounted-for systematics in the \planck\ data \cite{Motloch:2018pjy,DiValentino:2019qzk,Calderon:2023obf}. Fortunately, upcoming small-scale CMB observations from ACT and SPT, as well as galaxy power spectrum measurements by DESI \cite{DESI:2024jis,DESI:2024hhd}, will provide further insights into the physics at very high-energy scales.

\section{Conclusions}\label{sec:conclusion}

Probing the initial condition of the Universe is one of the main science goals for the next decade.  
While forthcoming CMB experiments will constrain the tensor-to-scalar ratio $r$ with unprecedented precision \cite{SimonsObservatory:2018koc,LiteBIRD:2022cnt,CMB-S4:2016ple,NASAPICO:2019thw,Sehgal:2019ewc}, clustering measurements of the large-scale structure offer a complementary window to study the inflationary epoch \cite{Chen:2016vvw,LSST,Amendola:2016saw,SPHEREx:2014bgr,Ferraro:2019uce,Ferraro:2022cmj,Sailer:2021yzm,DAmico:2022gki}.  
Crucially, LSS measurements provide access to a significantly larger number of modes compared to the CMB, enabling improved constraints on primordial features. This advantage will become even more pronounced with upcoming spectroscopic and photometric surveys capable of probing higher redshifts ($z \gtrsim 2-3$), where the number of observed modes increases substantially \cite{Ferraro:2019uce,Sailer:2021yzm}. As a result, these measurements will soon tighten constraints on the physics of the early universe, further refining our understanding of inflation and potential deviations from the standard slow-roll dynamics.

In this work, we applied the effective field theory of large-scale structure (EFTofLSS) to analyze the galaxy power spectrum multipoles measured by BOSS LRG and eBOSS QSO, in combination with CMB (T\&E) data, to constrain some representative primordial features that cover a wide range of inflationary dynamics.
A key strength of our approach is its ability to incorporate not only the non-linear effects from the long-wavelength modes (through IR-resummation) but also to account for small-scale non-linearities thanks to the one-loop EFT contributions, allowing us to analyze scales up to $k = 0.24 h {\rm Mpc}^{-1}$. In addition, we emphasize that our IR-resummation scheme does not rely on the wiggle/no-wiggle split procedure (as done in past literature~\cite{Beutler:2019ojk,Mergulhao:2023ukp,Ballardini:2022wzu}), but on an analytical formula (implemented in \pybird) which is (in principle) independent of the primordial feature scenario considered (see also \cite{Chen:2024pyp} for a similar LPT-based approach). 
We summarize our results as follows:
\begin{itemize}
    \item  We first validated the EFTofLSS predictions for the matter power spectrum (including the IR-resummation) against the results from N-body simulations for both linear and logarithmic (global) primordial oscillations in \cref{sec:Nbody}. 
    In both cases, we find a very good ($<0.5\%$) agreement with the semi-analytical fitting formulae derived from N-body simulations in Ref.~\cite{Ballardini:2019tuc}, enabling us to check the good accuracy of the IR-resummation scheme implemented in \pybird. This is illustrated in \cref{fig:comparison_Nbody_IR}. 
    
    \item When applied to real data, we find that a full-shape analysis of current LSS data can already place tight constraints on various early universe models. 
    In \cref{sec:Global}, we focus on global oscillatory features, where we obtain strong constraints on the amplitudes of the linear and logarithmic primordial oscillations, with $A_{\rm lin}^{\rm LSS} < 0.031$ and $A_{\rm log}^{\rm LSS} < 0.024$ (at $95\%$ CL). When we combine the LSS analysis with the primary CMB power spectra from \planck, we obtain $A_{\rm lin}^{\rm LSS + CMB} < 0.022$ and $A_{\rm log}^{\rm LSS + CMB} < 0.021$ (at $95\%$ CL), which significantly improves the CMB constraints by $\sim 25 \%$ and $\sim 50 \%$, respectively.
    
    \item We then applied this analysis to local features scenarios (taking place in the scales probed by our LSS dataset), showing that an EFTofLSS analysis applied to (e)BOSS data can strongly disfavor localized primordial oscillations that are favored by \planck. 
    In the case of the \OSpec template discussed in \cref{sec:one_spec}, we constrain the amplitude of the features to be $\alpha < 0.094$ (at $95\%$ CL) using \planck~data alone, while our LSS analysis alone tightens this to $\alpha < 0.034$ (at $95\%$ CL), which is $\sim 3$ times better than the CMB constraint (with no CMB lensing).  
    Our validation tests with synthetic data reveal that localized features like those induced by the \OSpec template with large amplitude ($\alpha=0.1$)\textemdash that are still allowed by \planck \textemdash would be detected by the LSS data, as illustrated in \cref{fig:OneSpectrum_mock}.
    This highlights the enhanced sensitivity of LSS data to such features, particularly in the wavenumber range $0.14 \lesssim k \lesssim 0.24 ~h{\rm Mpc}^{-1}$, where the amplitude of oscillations reaches its maximum (see \cref{fig:OS_mPk_residuals}).
  
    \item Interestingly, while nonlinear structure formation processes tend to wash out oscillatory features—especially at low redshifts—we find that primordial features with sufficiently large amplitudes can survive and leave detectable imprints in current and upcoming LSS measurements \cite{Sailer:2021yzm,Ferraro:2022cmj}. 
    
    \item Finally, in \cref{sec:regMRL}, we applied our methodology to free-form reconstructions of the PPS using CMB temperature and polarization anisotropies. Interestingly, while the deconvolved PPS provides a better fit to \planck\ TT, TE, and EE data simultaneously~\cite{Sohn:2022jsm}, the reconstructed features are disfavored by LSS data.  
    The lack of overlap between the CMB and LSS contours in \cref{fig:MRL} demonstrates that an EFT-based analysis of LSS data can place strong constraints on a PPS reconstructed from CMB data. 
    This underscores the importance of cross-validating PPS reconstructions from CMB and LSS data to ensure consistency across different cosmological probes.  
\end{itemize}

There are several ways in which our work can be improved and extended.  
First, the frequency range of the global primordial features can be expanded to include higher frequencies by reducing the bandwidth $\Delta k$, following the approach in Refs.~\cite{Beutler:2019ojk,Mergulhao:2023ukp}.  
Second, our analysis can be extended beyond two-point statistics by incorporating bispectrum measurements, as recently developed in Refs.~\cite{DAmico:2022ukl,Spaar:2023his,Chen:2024pyp}, which provide additional sensitivity to primordial features.  
Third, a joint reconstruction of the PPS using both CMB and LSS data could be performed, for instance, by applying the regularized modified Richardson-Lucy methodology simultaneously to both datasets.  
Finally, our analysis pipeline is readily applicable to power spectrum measurements from DESI~\cite{DESI:2024hhd} and the forthcoming \textit{Euclid} mission~\cite{EUCLID:2011zbd,Euclid:2023shr}, which will provide higher-precision data and further refine constraints on primordial features.

\acknowledgments
We thank Xingang Chen, Guido D'Amico, Fabio Finelli, Thiago Mergulh\~ao, Vivian Poulin, and Pierre Zhang for insightful feedback and comments at various stages of the project, as well as Wuhyun Sohn for interesting discussions and for sharing the regMRL reconstructed PPS.
We would also like to acknowledge all the participants to the \href{https://indico.gssi.it/event/606/}{ModIC 2024 workshop} for the fruitful interdisciplinary exchange of ideas. R.C. is funded by the Czech Ministry of Education, Youth and Sports (MEYS) and European Structural and Investment Funds (ESIF) under project number CZ.02.01.01/00/22\_008/0004632. A.S. would like to acknowledge the support by National Research Foundation of Korea 2021M3F7A1082056, and the support of the Korea Institute for Advanced Study (KIAS) grant funded by the government of Korea. DKH acknowledges India-Italy ``RELIC - Reconstructing Early and Late events In Cosmology" mobility program and the Indo-French Centre for the Promotion of Advanced Research – CEFIPRA grant no. 6704-4. 
TS acknowledge the European Union’s Horizon Europe research and innovation programme under the Marie Skłodowska-Curie Staff Exchange grant agreement No 101086085 – ASYMMETRY.
This work was supported by the computational resources from the  LUPM’s cloud computing infrastructure founded by Ocevu labex and France-Grilles, as well as the high-performance computing cluster Seondeok at the Korea Astronomy and Space Science Institute.
We acknowledge the use of the Boltzmann solver \classy~\cite{class_2011arXiv1104.2932L,class_2011JCAP...07..034B} for the computation of theoretical observables, 
\texttt{Montepython}\ \citep{Brinckmann:2018cvx,Audren:2012wb} for the sampling and \gd\ \cite{Lewis:2019xzd} for the post-processing of our results. We also acknowledge the use of the standard \texttt{python} libraries for scientific computing, such as \texttt{numpy} \cite{harris2020array}, \texttt{scipy} \cite{2020SciPy-NMeth} and \texttt{matplotlib} \cite{Hunter:2007}.

\appendix

\section{Supplementary material: Linear and logarithmic oscillations} \label{app:lin_log}

\begin{figure}
    \centering
    \includegraphics[width=\linewidth]{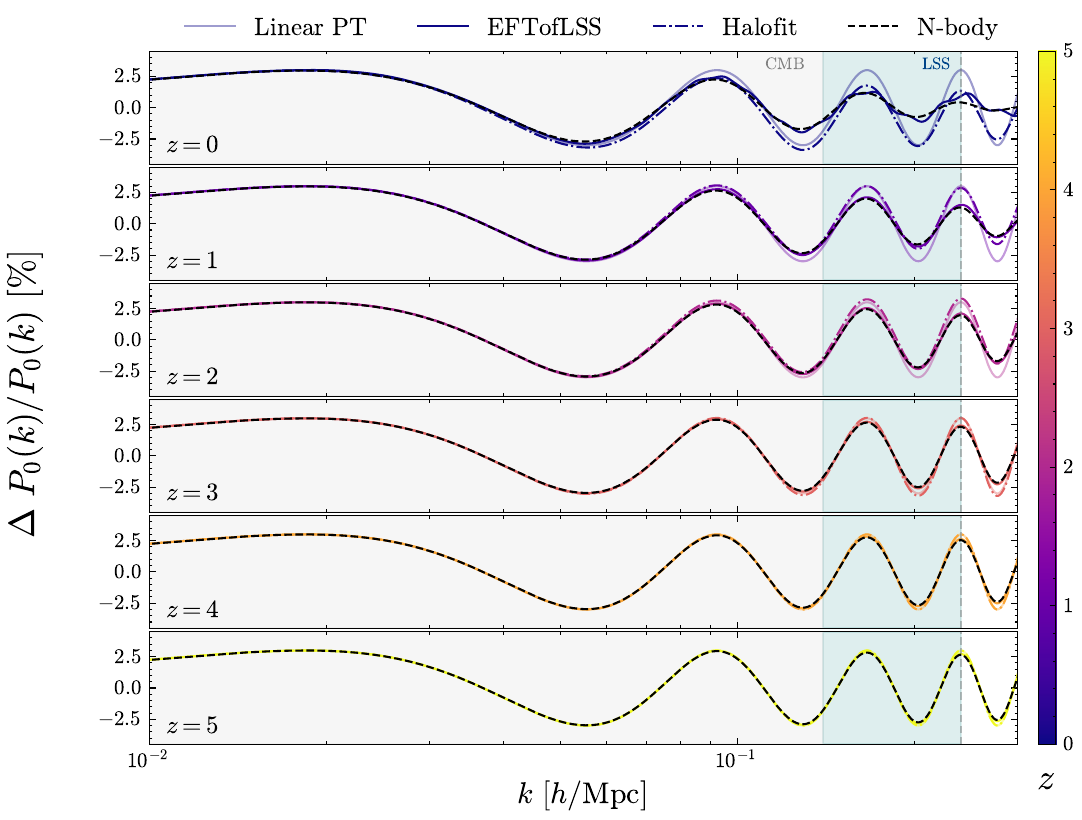}
    \caption{Non-linear matter power spectrum residuals of the linear oscillations (see \cref{eq:log-osc}) with respect to a featureless power-law, considering an amplitude $A_{\rm lin}=0.03$, a frequency $\log_{10}\omega_{\rm lin}=0.8$, and a phase $\phi_{\rm lin}=0$. 
    We show the prediction from the linear perturbation theory,  the EFTofLSS, \texttt{Halofit}, as well as from the analytical fitting formulae of \cite{Ballardini:2019tuc}, based on N-body simulations.
    This figure shows an excellent between the EFTofLSS prediction and the N-body simulations.}
    \label{fig:Nbody-comp-lin}
\end{figure}

\begin{figure}
    \centering
    \includegraphics[width=0.49\textwidth]{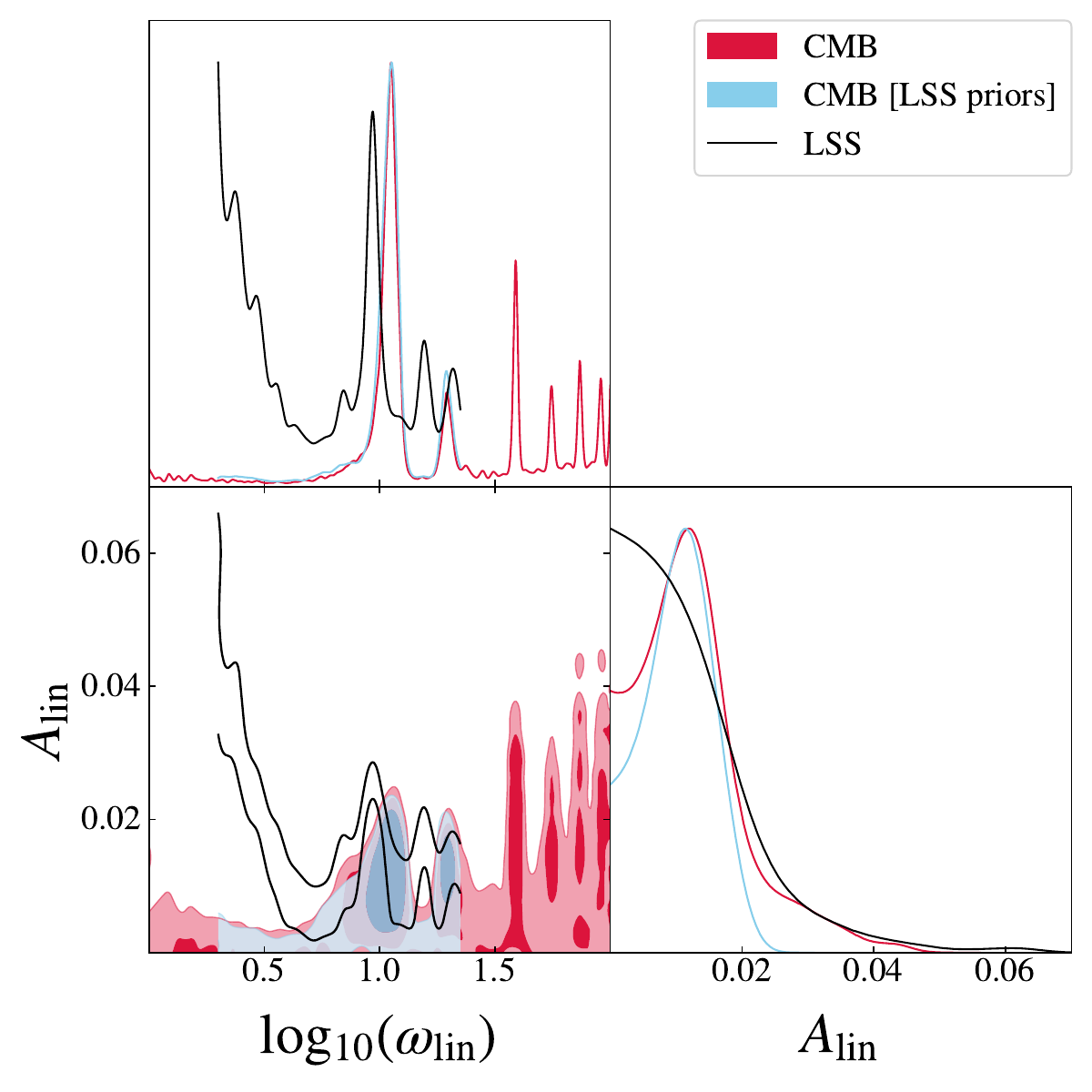}
    \includegraphics[width=0.49\textwidth]{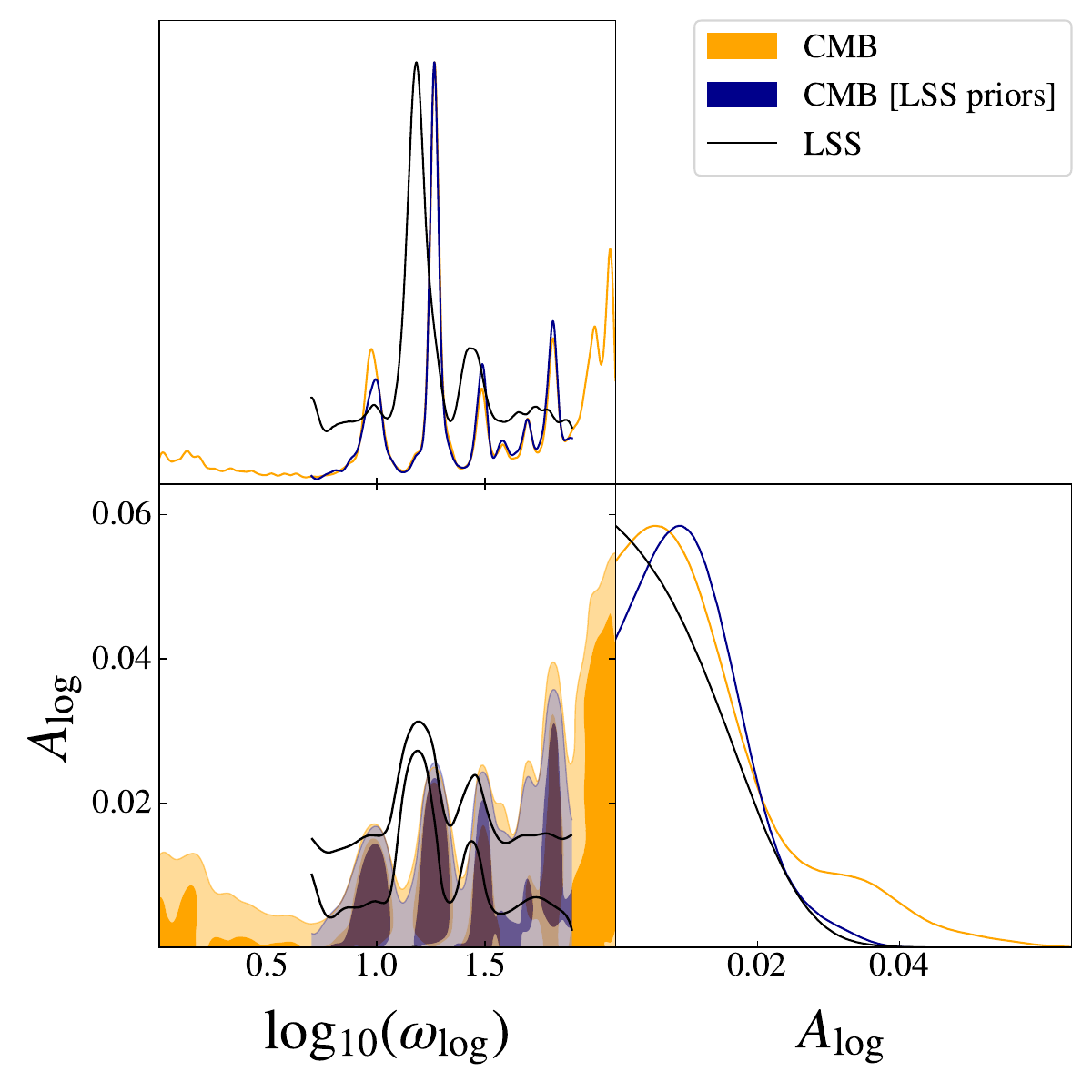}
    \caption{\textit{Left:} 1D and 2D posterior distributions of the $\log_{10}(\omega_{\rm lin})- A_{\rm lin}$ plan reconstructed from the CMB, CMB with the LSS prior on $\log_{10}(\omega_{\rm lin})$, and LSS analyses for the linear oscillations. \textit{Right:} Same for the logarithmic oscillations.}
    \label{fig:w_A_LSS_priors}
\end{figure}

\begin{figure}
    \centering
    \includegraphics[width=0.49\columnwidth]{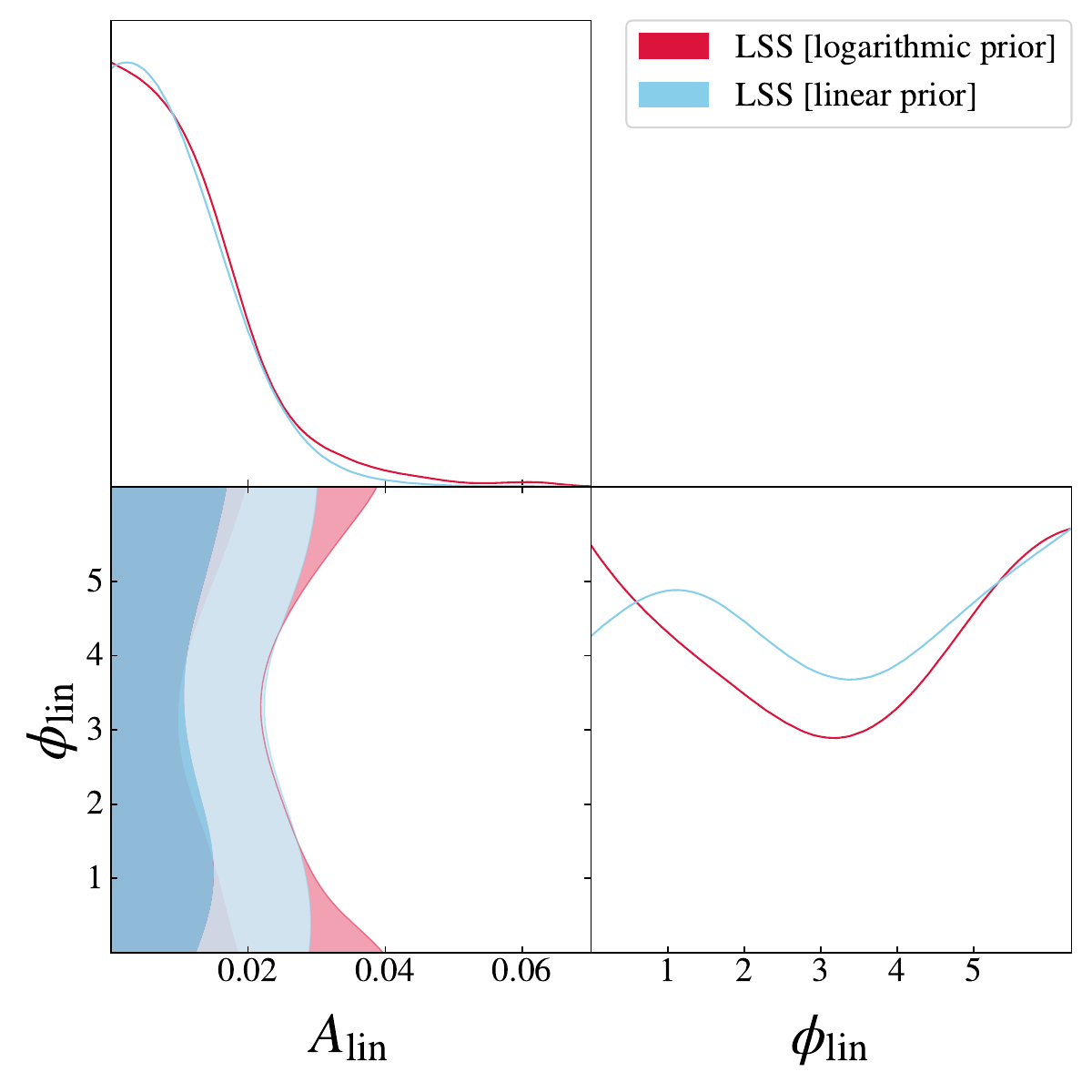}
    \includegraphics[width=0.49\columnwidth]{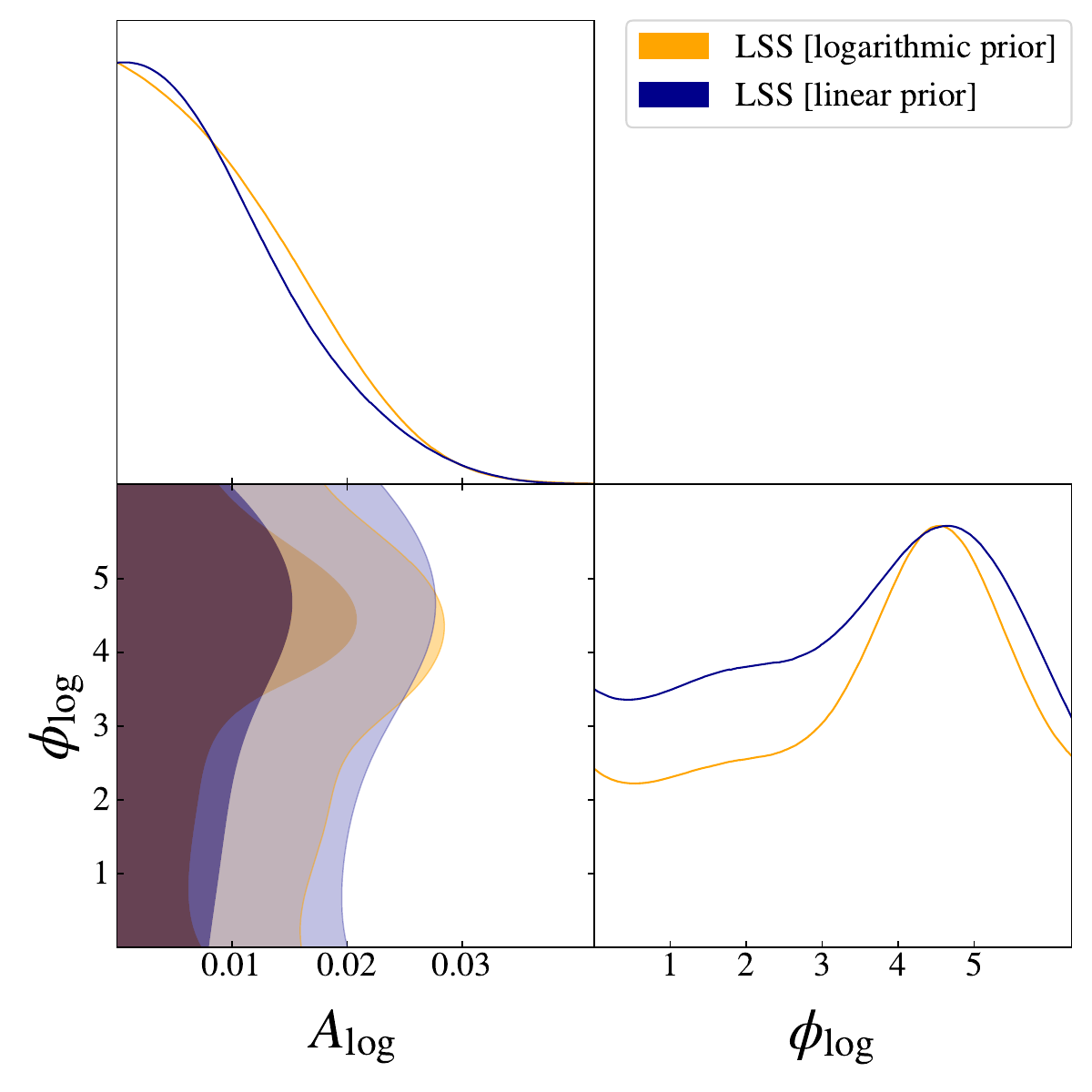}
    \caption{\textit{Left:} 1D and 2D posterior distributions of $A_{\rm lin} - \phi_{\rm lin}$ reconstructed from the LSS dataset with the logarithmic and the linear priors on $\omega_{\rm lin}$ for the linear oscillation analysis. \textit{Right:} Same for the logarithmic oscillation analysis.}
    \label{fig:lin_prior_lin_log}
\end{figure}

In this appendix, we show additional figures for linear and logarithmic oscillations:
\begin{itemize}
    \item In \cref{fig:Nbody-comp-lin}, we show the equivalent of \cref{fig:Nbody-comp} for the linear oscillations, corresponding to the matter power spectrum residuals (with respect to $\Lambda$CDM), as a function of redshift, inferred from linear perturbation theory, EFTofLSS, \texttt{Halofit}, and the N-body simulation of Ref.~\cite{Ballardini:2019tuc}. The conclusions are the same as the logarithmic oscillations (see \cref{fig:Nbody-comp}).
    \item In \cref{fig:w_A_LSS_priors}, we display, for  linear and logarithmic oscillations, the $\log_{10}(\omega_{\rm X})- A_{\rm X}$ plan reconstructed from the CMB, CMB with the LSS prior on $\log_{10}(\omega_{\rm lin})$, and LSS analyses. 
    This figure allows us to (i) compare the LSS and CMB constraints with the same prior on $\log_{10}(\omega_{\rm lin})$, and (ii) show that the CMB constraints with the restrictive prior are similar to the ones with the CMB prior. We further note that the restriction of this prior does not influence the other cosmological parameters.
    \item In \cref{fig:lin_prior_lin_log}, we compare the LSS constraints from the analysis with the logarithmic and the linear prior on the frequency $\omega_X$. We can see that the constraints on $A_X$ do not depend on the type of prior imposed on $\omega_X$ for both the linear and logarithmic oscillations.
\end{itemize}

\section{Supplementary material: One Spectrum-like template} \label{app:power_spectra_one_spectrum}

\begin{figure}
    \centering
    \includegraphics[trim=0.8cm 0cm 3cm 1.5cm, clip=true,width=1.\columnwidth]{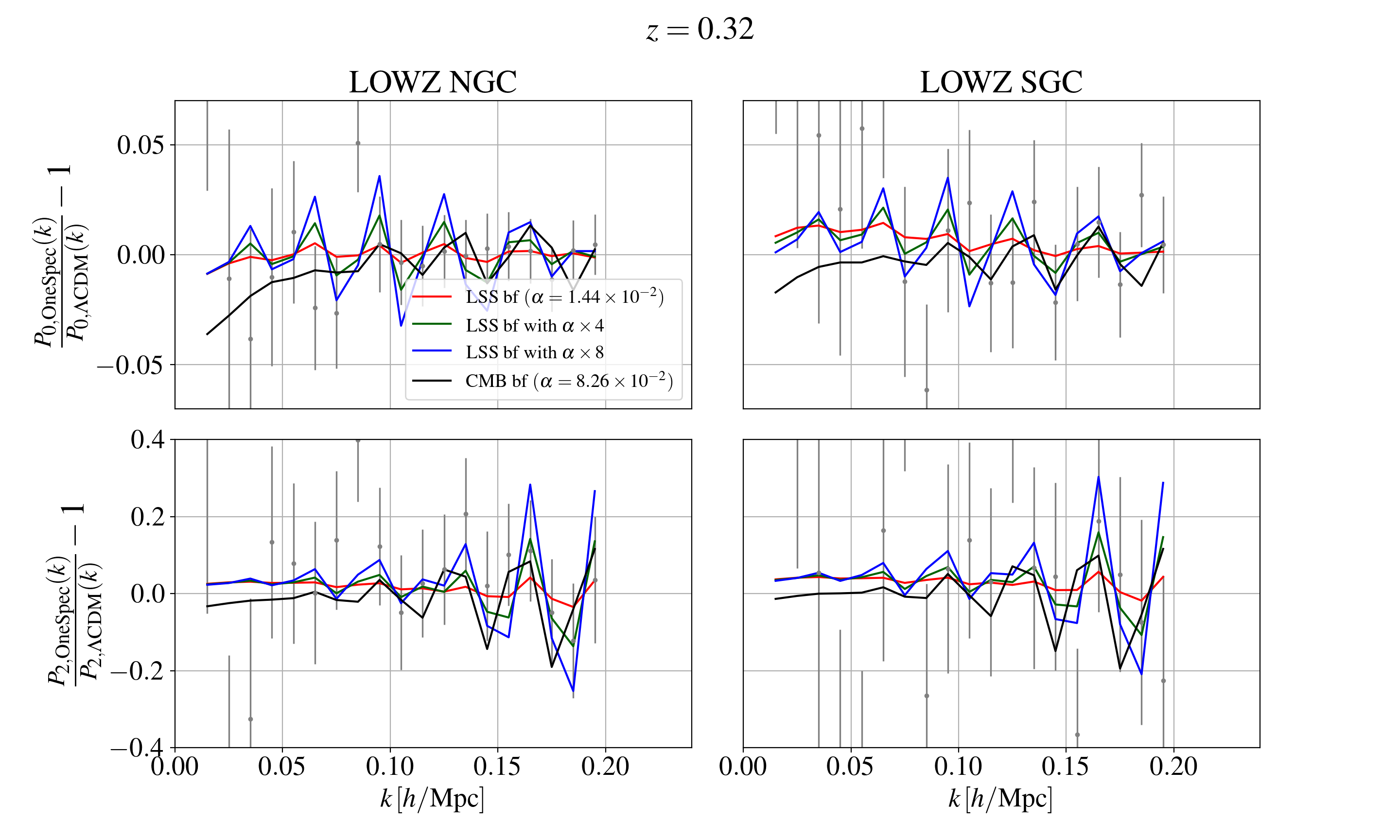}
    \includegraphics[trim=0.8cm 0cm 3cm 1.5cm, clip=true,width=1.\columnwidth]{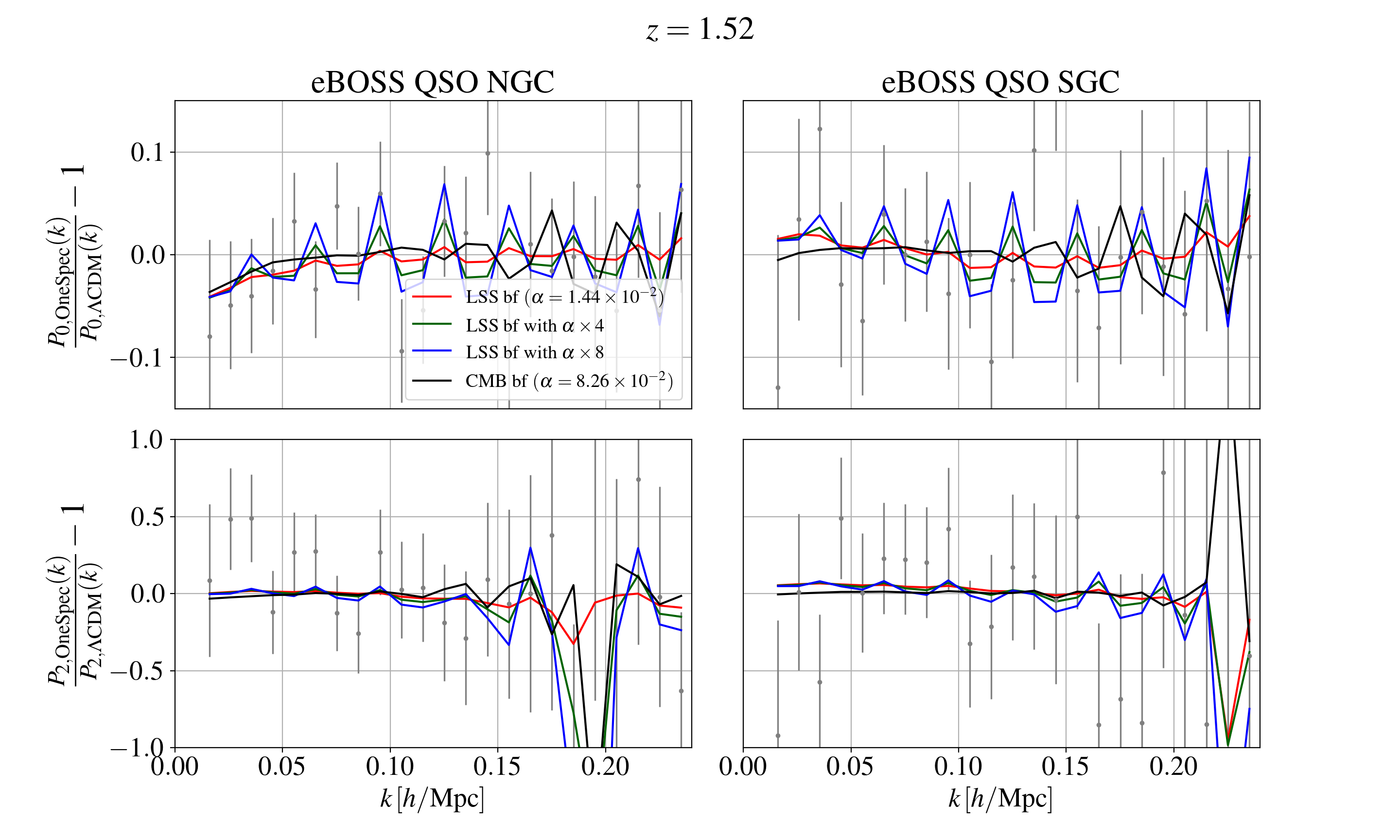}
    \caption{Same as \cref{fig:residuals_OneSpectrum_BOSS} for the two sky cuts of the BOSS LOWZ and eBOSS QSO data.}
    \label{fig:residuals_OneSpectrum_eBOSS}
\end{figure}

\begin{figure}
    \centering
    \includegraphics[width=1.\columnwidth]{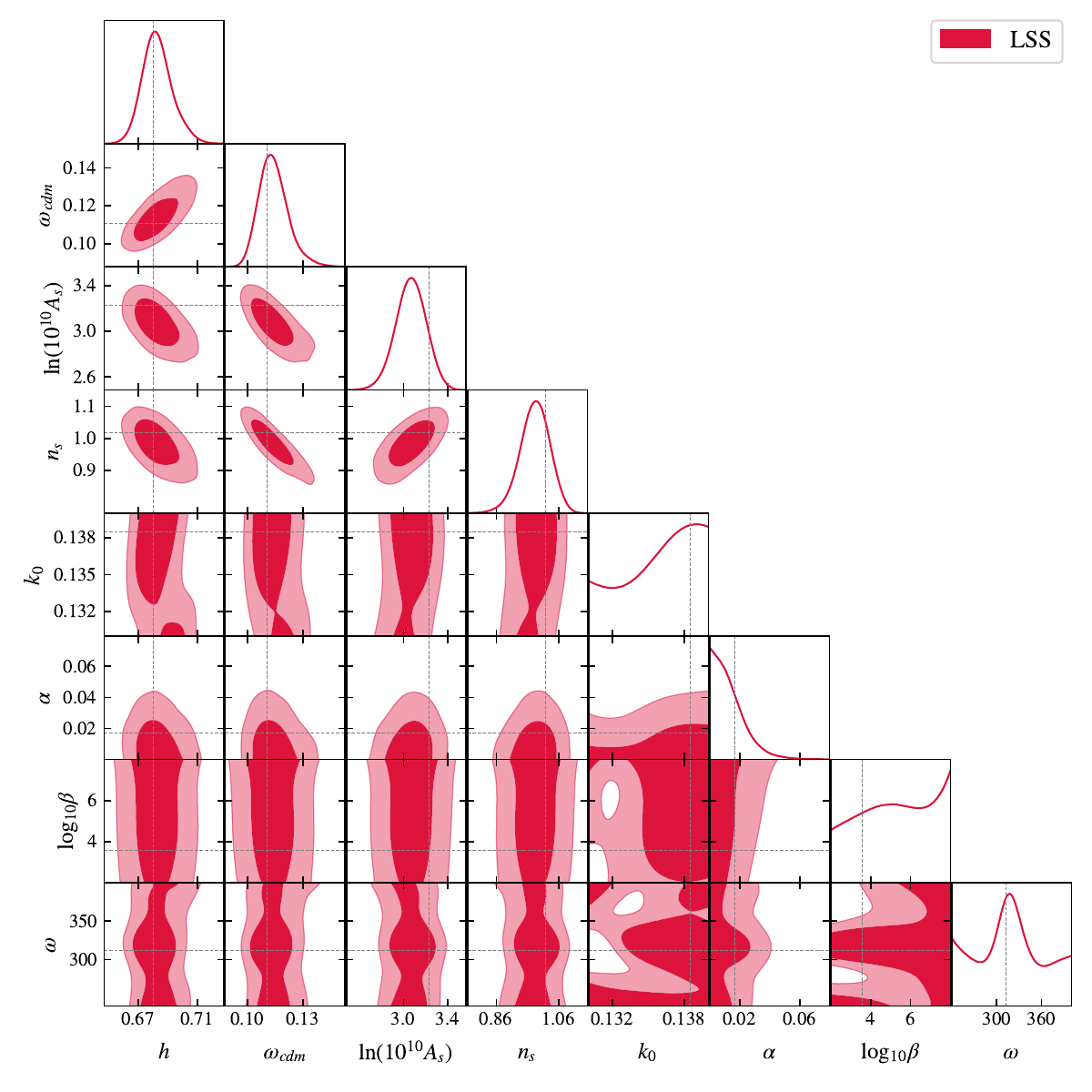}
    \caption{1D and 2D posterior distributions, together with the associated bestfit (in dashed lines), reconstructed from the LSS dataset for the \OSpec analysis.}
    \label{fig:One_spectrum_bestfit}
\end{figure}

In this appendix, we show additional figures for the \textit{One Spectrum} analysis:
\begin{itemize}
    \item In \cref{fig:residuals_OneSpectrum_eBOSS}, we show the residuals of the monopole and quadrupole of the \OSpec galaxy power spectra with respect to $\Lambda$CDM for the two sky cuts of the BOSS LOWZ and eBOSS QSO samples.
    \item In \cref{fig:One_spectrum_bestfit}, we display the 1D and 2D posterior distributions of the LSS analysis, together with the associated bestfit. We can see that the MAP values lie quite close to the median of the posterior distributions, suggesting that unlike many \lcdm\ extensions, this model does not suffer from strong projection  effects \cite{DESI:2024jis,DESI:2024hhd}.
\end{itemize}

\bibliographystyle{JHEP}
\bibliography{biblio.bib}

\end{document}